\documentclass[fleqn,usenatbib]{mnras}

\usepackage{newtxtext,newtxmath}

\usepackage{graphicx}	

\usepackage[T1]{fontenc}
\usepackage{times}
\usepackage{url}
\usepackage{graphicx}
\usepackage{graphics} 
\usepackage{epsfig}   
\usepackage{microtype}
\usepackage{threeparttable} 
\usepackage{pdflscape}	
\usepackage{natbib}
\usepackage{enumitem}
\setlist{leftmargin=5pt}

\def\aj{{AJ}}                   
\def\araa{{ARA\&A}}          
\def\apj{{ApJ}}                 
\def\apjl{{ApJ}}                
\def\apjs{ {ApJS}}               
\def\ao{ {Appl.~Opt.}}           
             
\def\aap{ {A\&A}}                
\def\aapr{ {A\&A~Rev.}}

\def\mnras{ {MNRAS}}

\def\pasp{ {PASP}}

\newcommand{\cxo}{{Chandra}}

\newcommand{\be}{\begin{equation}}
\newcommand{\ee}{\end{equation}}

\newcommand{\gtsima}{$\; \buildrel > \over \sim \;$}
\newcommand{\ltsima}{$\; \buildrel < \over \sim \;$}
\newcommand{\prosima}{$\; \buildrel \propto \over \sim \;$}
\newcommand{\gsim}{\lower.5ex\hbox{\gtsima}}
\newcommand{\lsim}{\lower.5ex\hbox{\ltsima}}
\newcommand{\simgt}{\lower.5ex\hbox{\gtsima}}
\newcommand{\simlt}{\lower.5ex\hbox{\ltsima}}
\newcommand{\simpr}{\lower.5ex\hbox{\prosima}}

\newcommand{\lx}{$L_{\rm X}$}
\newcommand{\lr}{$L_{\rm r}$}
\newcommand{\lxlr}{$L_{\rm X}/L_{\rm r}$}
\newcommand{\lxobslr}{$L_{\text{X,obs}}/L_{\rm r}$}

\newcommand{\hzrg}{{high-$z$ radio galaxy}}
\newcommand{\hzrgs}{{high-$z$ radio galaxies}}
\newcommand{\beq}{$B_{\text{eq}}$}
\newcommand{\lxcmb}{{$L_{\text{X,CMB}}$}}
\newcommand{\lxssc}{{$L_{\text{X,SSC}}$}}
\newcommand{\lxobs}{{$L_{\text{X,obs}}$}}

\usepackage{graphicx}	
\usepackage{amsmath}	
\usepackage{caption}

\usepackage{siunitx}

\title[IC/CMB in \hzrgs]{Proof of CMB-driven X-ray brightening of high-$z$ radio galaxies}
\author[Hodges-Kluck et al.]
{Edmund~Hodges-Kluck$^{1}$,
Elena~Gallo$^2$,
Gabriele~Ghisellini$^3$,
Francesco~Haardt$^{4,5}$,\newauthor
Jianfeng~Wu$^6$,
Benedetta~Ciardi$^7$ \\\\
$^1$ NASA/GSFC, Code 662, Greenbelt, MD 20771, USA\\
$^2$ Department of Astronomy, University of Michigan, 1085 S University Ave, Ann Arbor, MI 48109, USA \\
$^3$ INAF -- Osservatorio Astronomico di Brera, Via Bianchi 46, I-23807 Merate, Italy\\
$^4$ Universit\'a degli Studi dell'Insubria, Via Valleggio 11, I-22100 Como, Italy \\
$^5$ INFN, Sezione Milano-Bicocca, P.za della Scienza 3, I-20126 Milano, Italy\\
$^6$ Department of Astronomy, Xiamen University, Xiamen, Fujian 361005, China\\
$^7$ Max Planck Institute for Astrophysics, Karl-Schwarzschild-Strasse 1, D-85748 Garching bei M{\"u}nchen, Germany\\
  }

\begin{document}
\maketitle
\begin{abstract}
We present a definitive assessment of the role of Inverse Compton scattering of Cosmic Microwave Background photons (IC/CMB) in the context of radio galaxies.  Owing to the steep increase of the CMB radiation energy density, IC/CMB is supposed to become progressively more important with respect to radio synchrotron cooling as the redshift increases. For typical energies at play, this process will up-scatter the CMB photons into the X-ray band, and is thus expected to yield a redshift-dependent, concurrent X-ray brightening and radio dimming of the jet-powered structures. Here we show how a conclusive proof of this effect hinges on high-resolution imaging data in which the extended lobes can be distinguished from the compact hot spots where synchrotron-self-Compton dominates the X-ray emission regardless of redshift. We analyze Chandra and Very Large Array data of 11 radio galaxies between $1.3\simlt z \simlt 4.3$, and demonstrate that the emission from their lobes is fully consistent with the expectations from IC/CMB in equipartition. Once the dependence on size and radio luminosity are properly accounted for, the measured lobe X-ray luminosities bear the characteristic $\propto (1+z)^4$ proportionality expected of a CMB seed radiation field. Whereas this effect can effectively quench the (rest-frame) GHz radio emission from $z\simgt 3$ radio galaxies below $\simlt$1 mJy, IC/CMB alone can not be responsible for a deficit in high-$z$, radio-loud AGN if--as we argue--such AGN typically have bright, compact hot spots.
\end{abstract}


\section{Introduction}
\label{section:intro}

Radio emission from active galactic nuclei (AGN) is typically associated with magnetized jets of relativistic, charged particles launched very close to the central engine that produce synchrotron emission. These jets are important not just to understanding the AGN phenomenon, but also as agents of AGN ``feedback'' in galaxies \citep[e.g.,][]{Moster2010} and galaxy clusters \citep{Fabian2012}, where AGN energy is needed to (re-)heat or expel gas to prevent much more rapid star formation than is observed. In a minority of AGN--including radio galaxies, radio-loud quasars and blazars--the radio luminosity substantially exceeds the optical luminosity; these are usually referred to as radio-loud AGN. Although the most radio-luminous AGN are canonically associated with very massive black holes, core radio emission from compact jets appears ubiquitous across the black hole mass function, including in low-luminosity AGN \citep{Nagar2005} and dwarf galaxies \citep{Reines20}. By analogy with Galactic black-hole X-ray binaries, the presence (and to some extent the luminosity) of a collimated jet depends strongly on the ratio of the mass accretion rate to the Eddington limit \citep{fbg04}. Thus, the frequency of radio-loud AGN as a function of redshift could be sensitive to the formation mechanism and evolution of massive black holes over cosmic time, as well as their impact on the galaxies or clusters that occupy the same dark matter halo. 

Below $z\simlt 2$, the fraction of radio-loud quasars is 10-20\% \citep{Kellermann1989,Padovani1993}. There is some evidence that the radio-loud fraction evolves with redshift \citep[e.g.,][]{Jiang2007}, but this remains a matter of debate \citep[e.g.,][]{Banados2015}. One confounding effect is that the expected observability of even the brightest, intrinsically radio-loud AGN at higher redshifts is questionable \citep{ajello09,volonteri11,ghis_cel_tav14}. This is because the relativistic particles that emit radio synchrotron emission can also cool via Inverse Compton (IC) scattering, in which an ambient or external photon field is boosted by interaction with higher-energy particles and leaves the system. Since the energy density of the CMB radiation scales as $u_{\rm CMB} \propto (1+z)^4$, cooling from IC scattering of the Cosmic Microwave Background photons (hereafter IC/CMB) is expected to become increasingly important at progressively higher redshifts, and eventually overtake synchrotron cooling. 
IC scattering boosts photons of initial frequency $\nu_0$ to $\nu_1 \approx 4\nu_0 \gamma^2/3$, where $\gamma$ is the relativistic particle Lorentz factor. Since $\nu_{\rm CMB} = 160.4(1+z)$~GHz, CMB photons will be scattered in the X-ray band for $\gamma \gtrsim 700(1+z)^{-1/2}$. Under the usual assumption that the jet particle energies are distributed as a power law with $N(\gamma) \propto \gamma^{-p}$, and with typical bounds of $10 < \gamma < 10^4$ \citep{Worrall2006}, then a significant fraction of the total IC/CMB luminosity will land in the (rest-frame) X-ray band. Thus, in the context of jetted AGN, IC/CMB is expected to produce X-rays from the same regions where the radio synchrotron emission is usually observed.

This mechanism will compete with synchrotron cooling in such a way that the ratio of the IC/CMB to synchrotron luminosity--which is well approximated by the \lx/\lr\ ratio--is equal to the ratio of the CMB to local magnetic energy density, $u_{\rm CMB}/u_{\rm B}$. Assuming that radio-loud AGN have a somewhat well-defined, redshift independent, average $u_{\rm B}$, then one expects that the ratio \lxlr\ increases as $(1+z)^4$. This implies a concurrent, redshift-dependent X-ray brightening and radio dimming of jetted AGN, so much so that these objects could possibly be disguised as radio-quiet AGN in spite of having powerful jets. This process is often referred to as CMB quenching \citep{ghisellini15}.

Although this expectation is marginally supported by the discovery of jets that are \textit{only} visible in the X-rays \citep{Simionescu2016,Schwartz2020}, so far there is no compelling evidence that, for radio-loud AGN, \lxlr\ is strongly correlated with redshift. \citet{smail12} and \citet{smail13} measured \lxlr\ in a sample of high-$z$ radio galaxies\footnote{Radio-loud AGN whose radio emission is resolved into separate components, possibly including a core, jet, ``hot spots,'' and diffuse lobes} with the Chandra X-ray Observatory and interferometric radio data, and found no correlation with redshift. Instead, \citet{smail13} noticed a weak correlation between \lxlr\ and the far-infrared luminosity, which led them to posit that IC scattering of local far-infrared photons (IC/FIR) dominates the IC X-ray emission instead. 
Likewise, fitting the spectral energy distributions of a few \hzrgs\ with a multi-component jet model, \citet{wu17} found evidence for X-ray IC/CMB emission consistent with a redshift enhancement in a few high-$z$ systems, but left open the possibility that many IC seed photons are FIR photons from the host. Meanwhile, \citet{zhu19} measured enhanced X-ray emission in highly radio-loud quasars at $z>4$, using a two-point spectral index analysis, but they concluded that the enhancement is too weak to be explained by a dominant role for IC/CMB at all redshifts (see also \citealt{miller11,wu13}). \citet{Ighina2019} found that the average \lxlr\ of radio-selected blazars at $z>4$ is higher than those at lower redshift, but without a strong correlation. Similar to \citet{zhu19}, they interpreted the enhancement as due to increased, but not dominant, IC/CMB.

In summary, whereas IC scattering of CMB photons off of relativistic AGN jet particles ought to (i) take place, so long as the CMB exists, and (ii) increase in relative strength with redshift, observational evidence for this effect is thin at best. In this paper, we demonstrate how the most likely explanation for this tension does not hinge on the presence of an additional seed photon field for IC. Rather, we show how prior studies have measured or inferred the \lx/\lr\ ratio from the {integrated emission} of the lobes plus the hot spots (and, occasionally, the core and jets as well). This, we argue, effectively washes out any possible redshift dependence, even for sources with similar lobe sizes. 

Throughout this work we adopt the following cosmology: $H_0 = 69.6$~km~s$^{-1}$~Mpc$^{-1}$, $\Omega_{\text{M}}=0.286$, and $\Omega_{\text{vac}} = 0.714$ \citep{Bennett2014}. 
\section{Inverse Compton scattering of CMB photons in equipartition} 

The lobes of powerful low-$z$ radio galaxies are typically close to equipartition between the magnetic field and relativistic particle energy density with $B \sim 0.3-1.3 B_{\text{eq}}$ and a strong peak near $0.7B_{\text{eq}}$ \citep{Croston2005,kataoka05}. In equipartition, $u_{\rm B} = B_{\rm eq}/8\pi$ can be estimated from the measured synchrotron luminosity, $L_{\rm sync}$, and emitting volume $V$ as:
\begin{equation}
    B_{\text{eq}} = \biggl[6\pi(1+k)c_{12} L_{\text{sync}} \phi^{-1} V^{-1}\biggr]^{2/7},
\end{equation}
where $k$ is the ratio of energy in protons to electrons, the function $c_{12}$ depends on the  frequency range and spectral index \citep{Pacholczyk1970}, and $\phi$ is the lobes' filling factor; the bulk of the synchrotron radiation is emitted in the radio band. Since we know how $u_{\rm CMB}$ evolves, we can then predict the luminosity arising from IC/CMB (as noted in \S~\ref{section:intro}, the Comptonized CMB photons will be boosted to X-ray energies for the typical parameters at play in radio galaxies). Notice that this is a lower bound to the actual expected luminosity because synchrotron self-Compton (SSC) could also produce X-rays. If IC/CMB is the only X-ray source, then the ratio of the X-ray luminosity expected from IC/CMB to the measured radio synchrotron luminosity can be written as:
\begin{equation}
    \frac{L_{\rm IC/CMB}} {L_{\rm{sync}}} = \frac{u_{\text{\rm CMB}}}{u_{\rm B}} \simeq  \frac{L_{\rm X,CMB}} {L_{\text{r}}}  = \frac{8\pi a T_0^4 (1+z)^4}{B^2},
\end{equation}
where $a = 7.5658\times 10^{-15}$~erg~cm$^{-3}$~K$^{-4}$ is the radiation constant and $T_0 = 2.725$~K is the temperature of the CMB at $z=0$. 
If we assume a spherical volume $(4/3)\pi r^3$, then equipartition implies that 
\begin{equation}
    L_{\rm X,CMB} = L_{\text{r}}^{3/7} \biggl[\frac{8\pi a T_0^4 (1+z)^4}{(\tfrac{9}{2}(1+k)c_{12})^{4/7}} (r^3 \phi)^{4/7}\biggr].
\end{equation}
We adopt $\phi = 1$, $k=0$ (electrons and positrons only), and $c_{12} = 5\times 10^7$. Assuming that the spectral index $\alpha \approx 0.7$ from 10~MHz up to 1-10~GHz, then $3 < c_{12}/10^7 < 10$. The X-ray band tends to probe electrons not yet affected by spectral ageing. This leads to:
\begin{equation}
    \biggl(\frac{L_{\rm X,CMB}}{10^{45} \text{ erg s}^{-1}}\biggr) = 2.6\times 10^{-5} \biggl(\frac{L_{\text{r}}}{10^{45} \text{ erg s}^{-1}}\biggr)^{3/7}  r_{\text{kpc}}^{12/7} (1+z)^4, 
    \label{eqn:beq}
\end{equation}
where $r_{\rm kpc}$ is the emitting region radius expressed in units of kpc.
If we can measure the size of the emitting region, and its radio luminosity, we can predict the IC/CMB X-ray luminosity in equipartition. We can then compare the predicted and measured X-ray luminosities to determine whether these assumptions are sufficient to explain the observations. Throughout this paper, we use rest-frame luminosities.

Here we perform this measurement for a small sample of \hzrgs\ with archival \cxo\ X-ray Telescope data. With its sub-arcsecond resolution, \cxo\ is necessary to resolve the X-ray emission from the lobes (vs. the hot spots), and to more closely match the resolution of the interferometric radio data. 
It is crucial to make this comparison in the lobes and hot spots \textit{separately}; albeit both may be in equipartition, Equation~\ref{eqn:beq} predicts a very different \lxlr\ ratio in the large, weakly magnetized lobes (higher \lxlr) as opposed to the small, strongly magnetized hot spots (lower \lxlr). In addition, SSC is likely to play a dominant role in the hot spots (Section~\ref{section:results}), enhancing their X-ray luminosity independent of redshift, so a clean and definitive test of whether the extended emission from radio galaxies is consistent with the expectations from IC/CMB in equipartition is to be carried out in the lobes. 

\begin{figure*}
    \centering
    \includegraphics[height=2in]{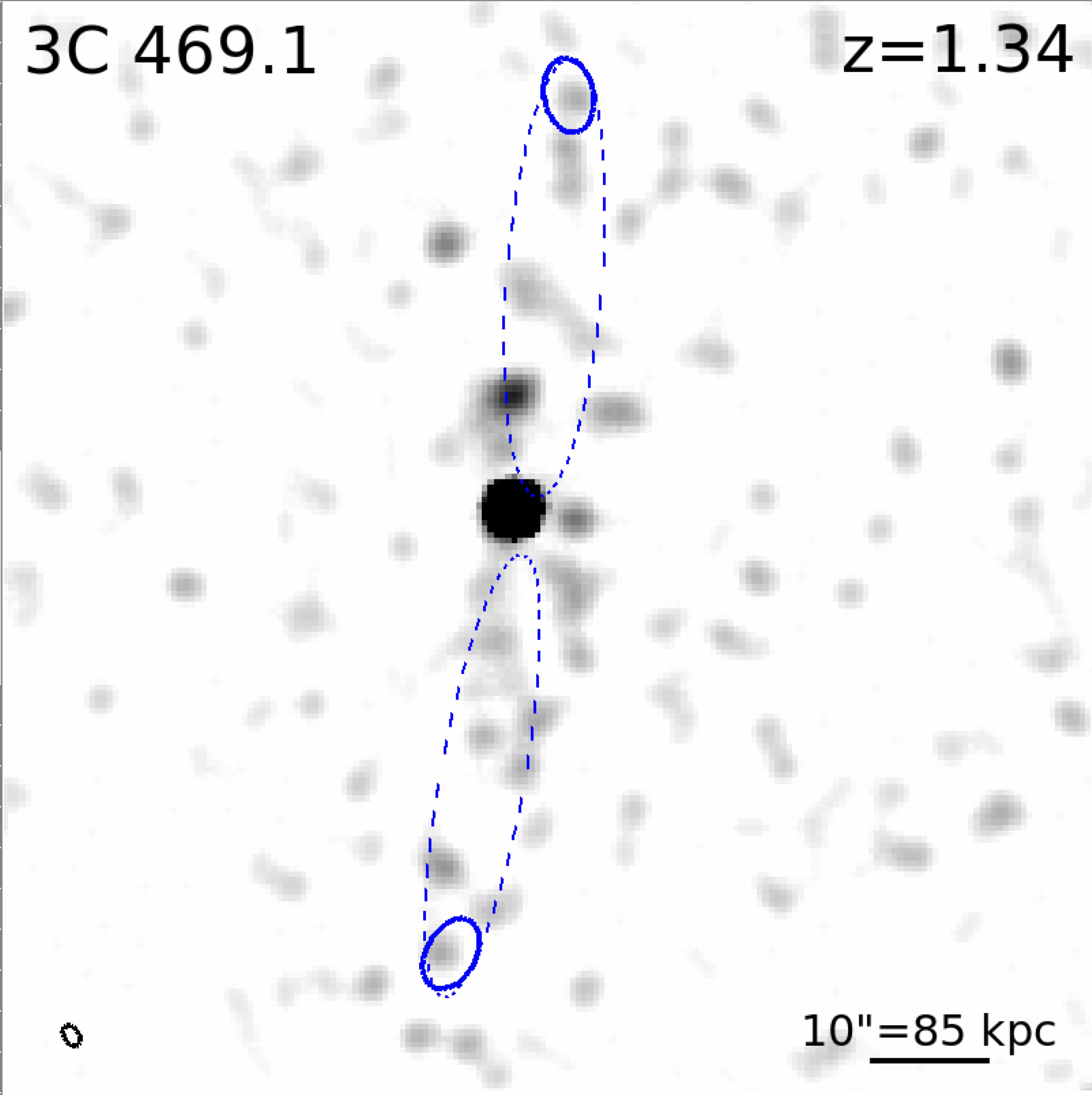}
    \includegraphics[height=2in]{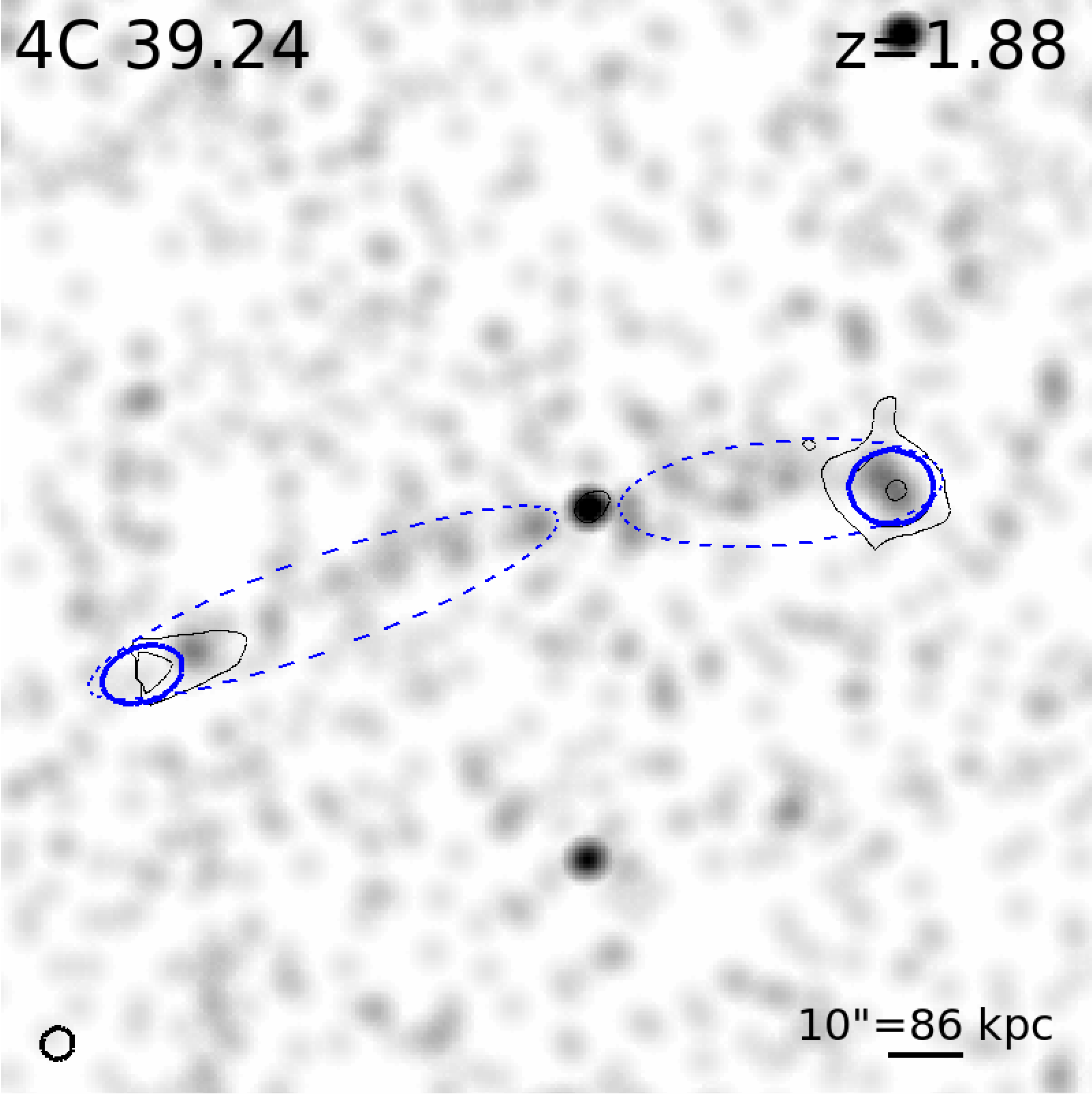}
    \includegraphics[height=2in]{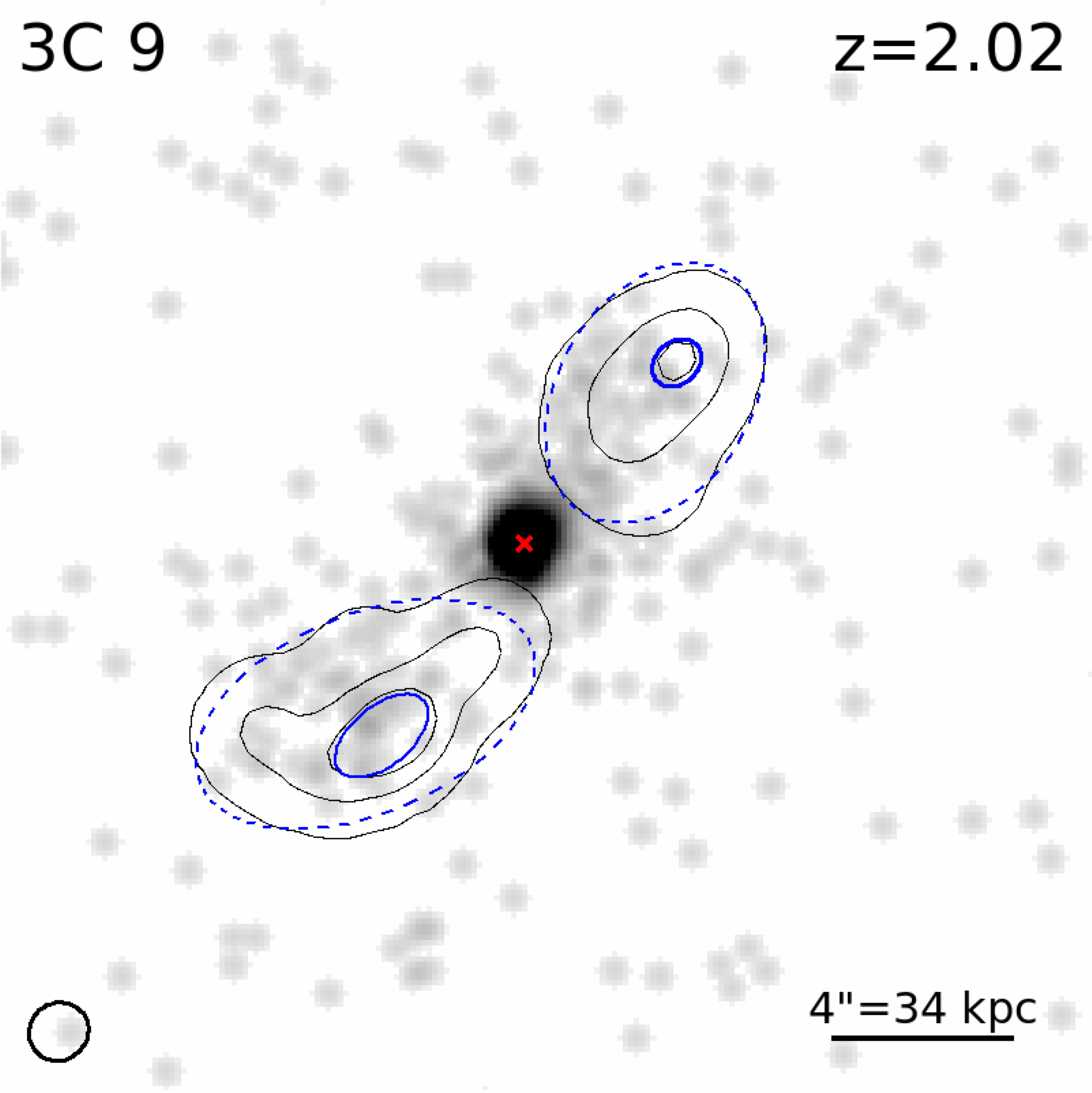}
    \includegraphics[height=2in]{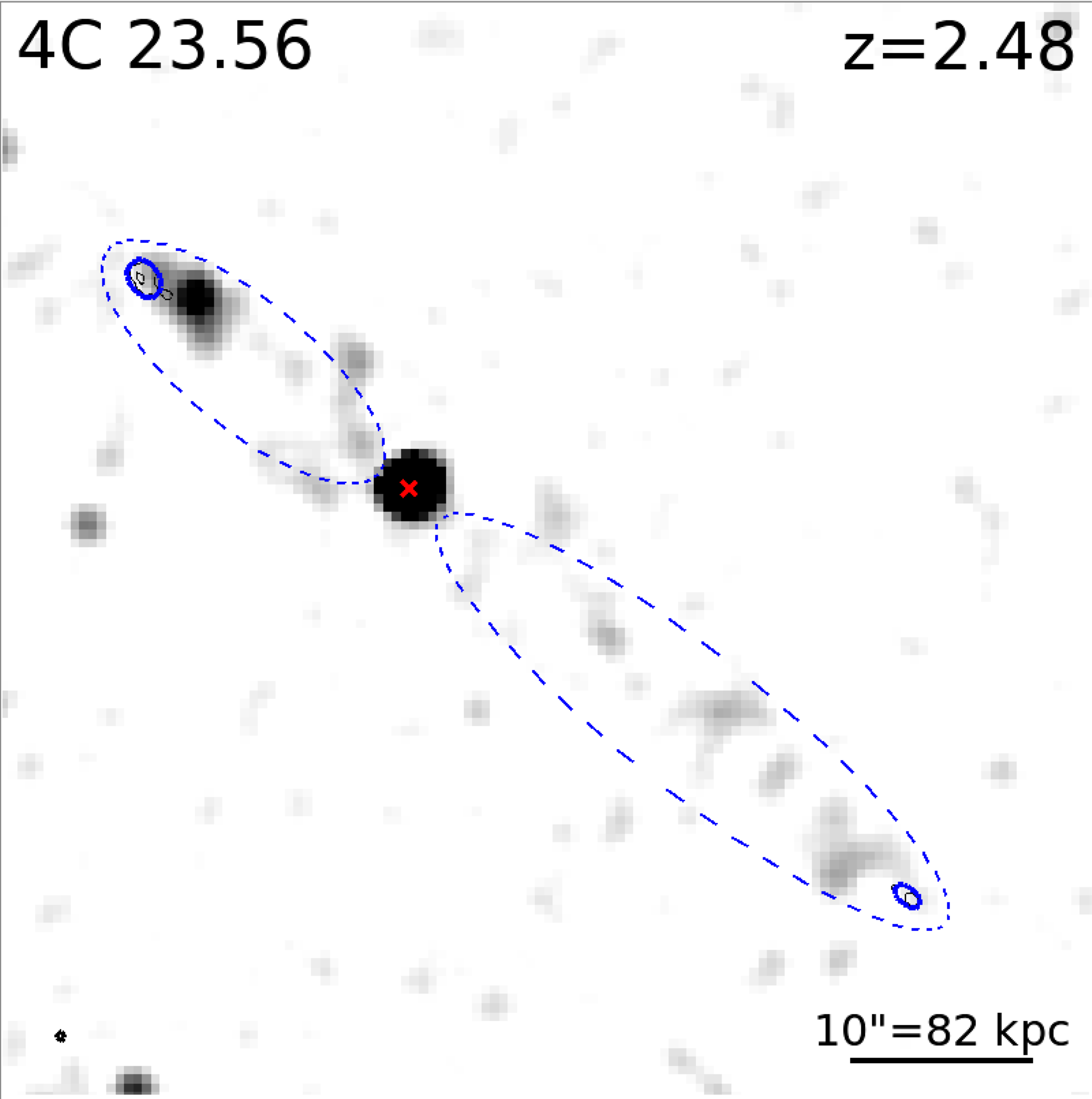}
    \includegraphics[height=2in]{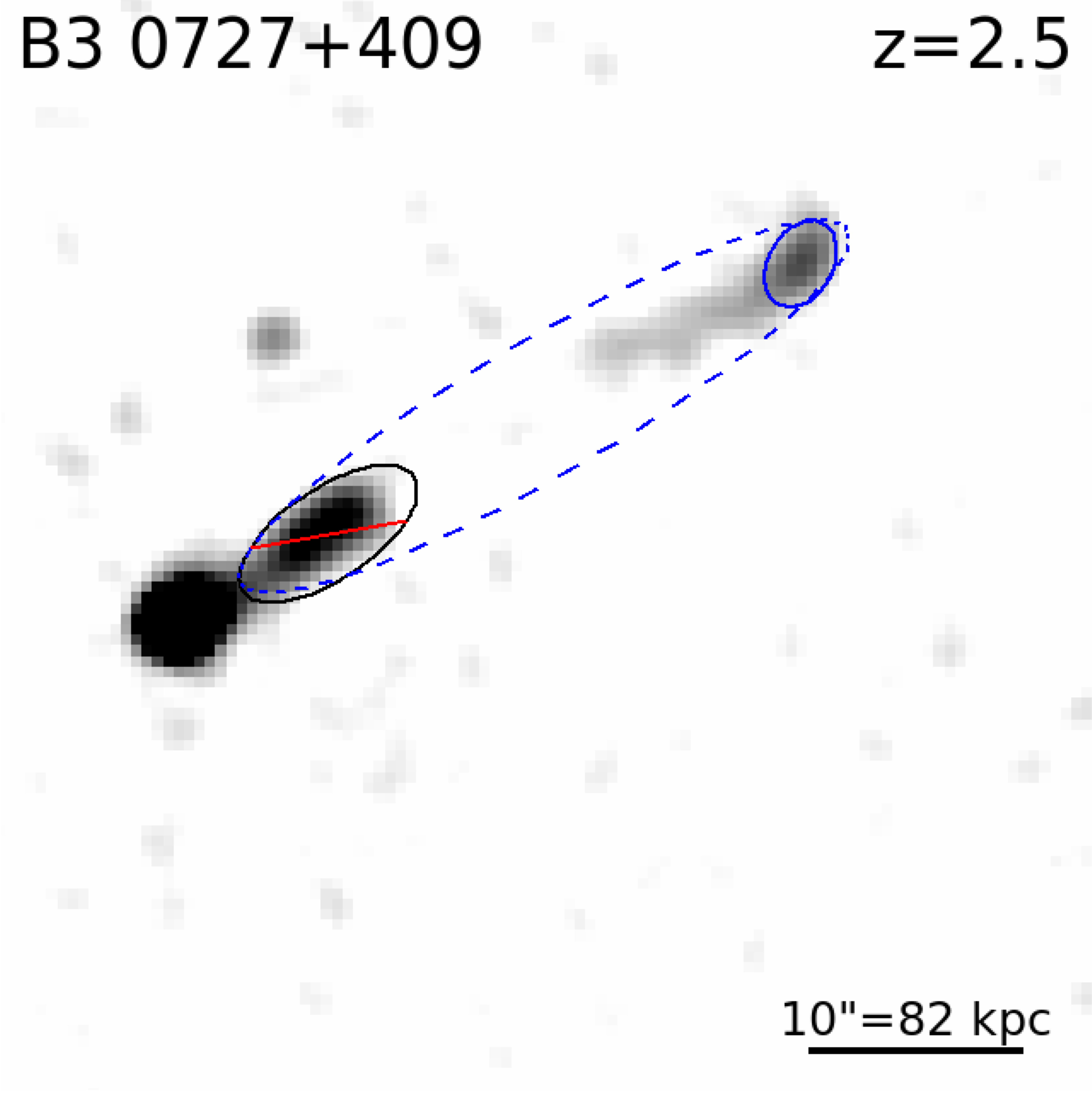}
    \includegraphics[height=2in]{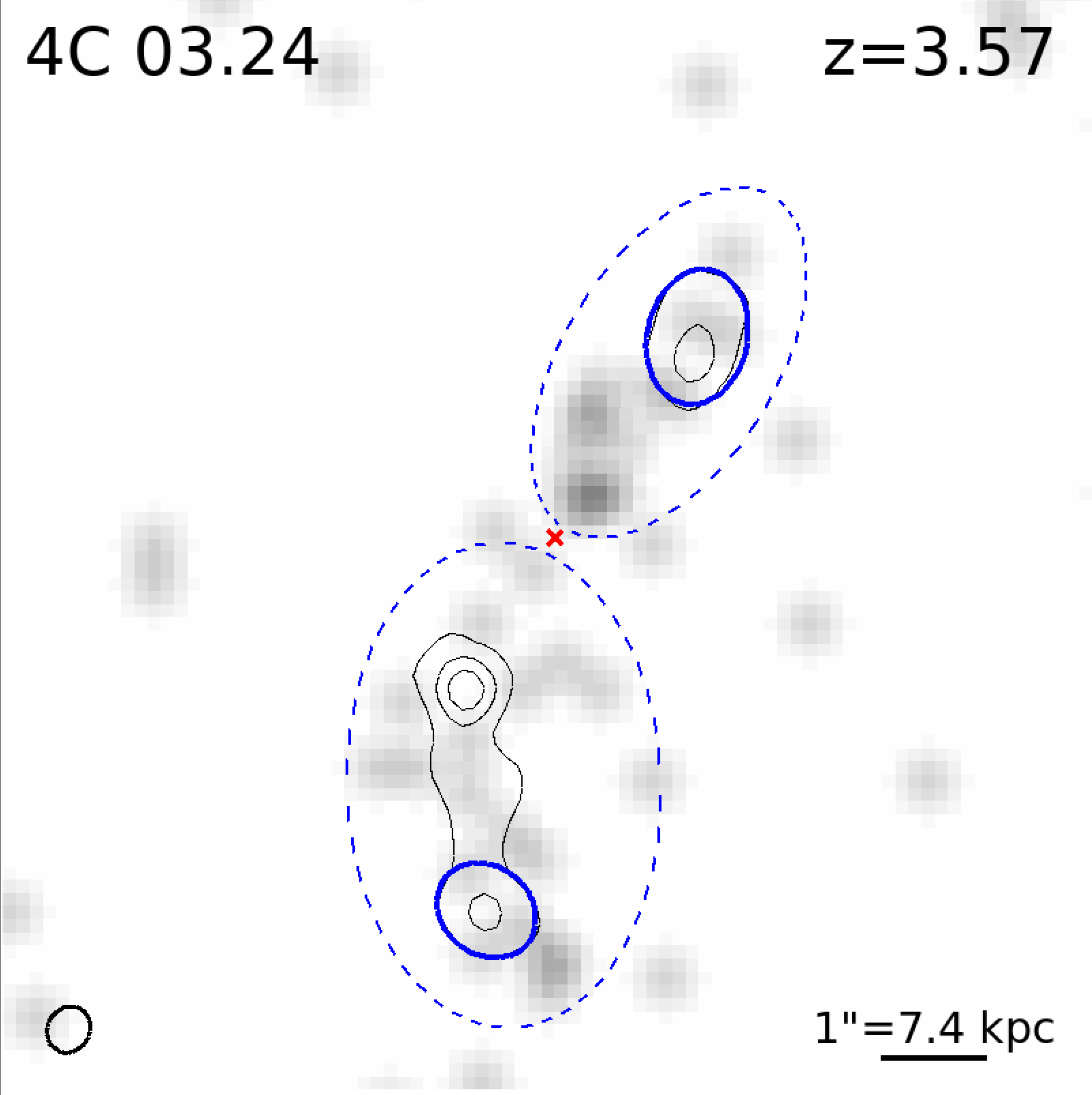}
    \includegraphics[height=2in]{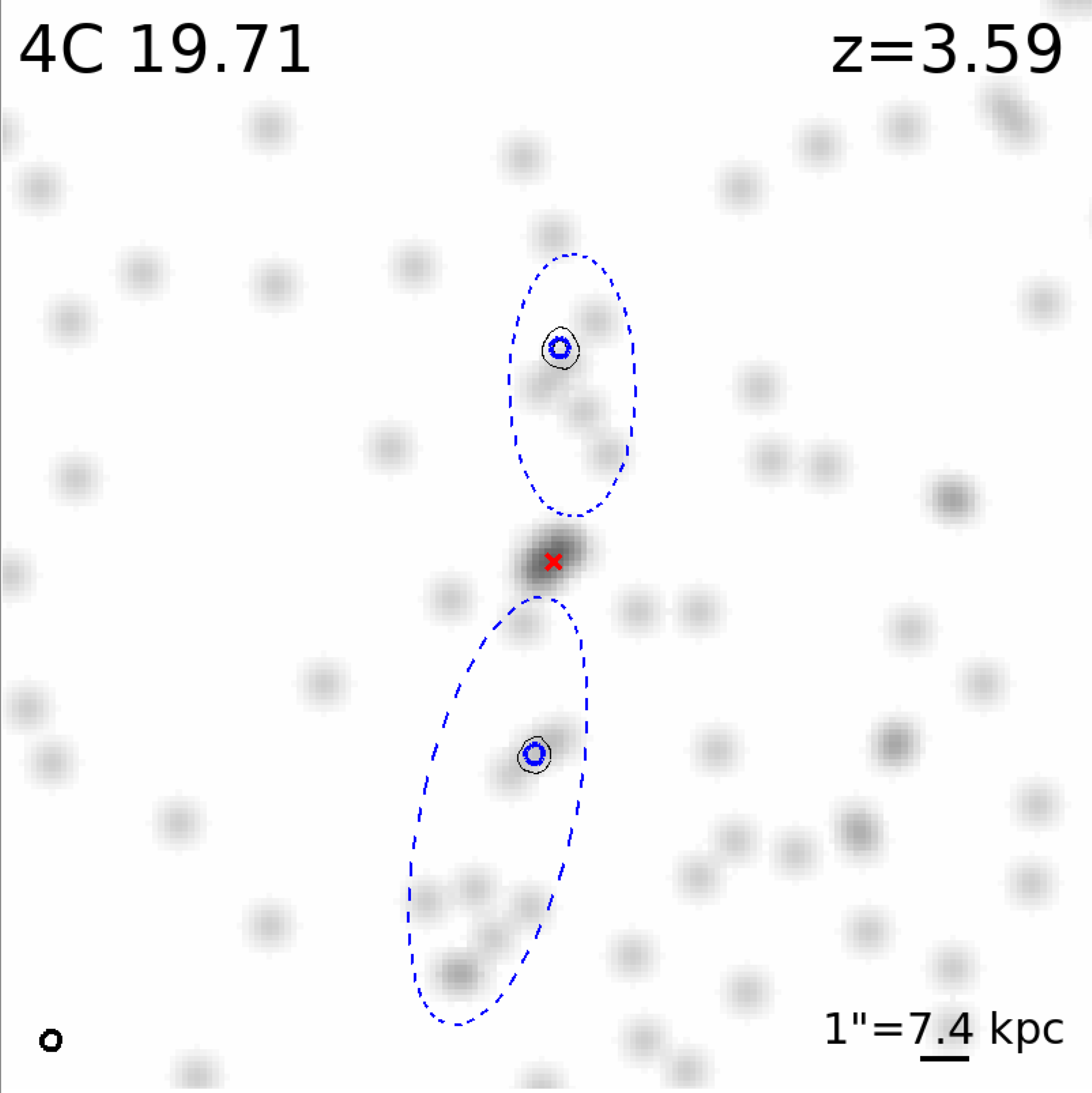}
    \includegraphics[height=2in]{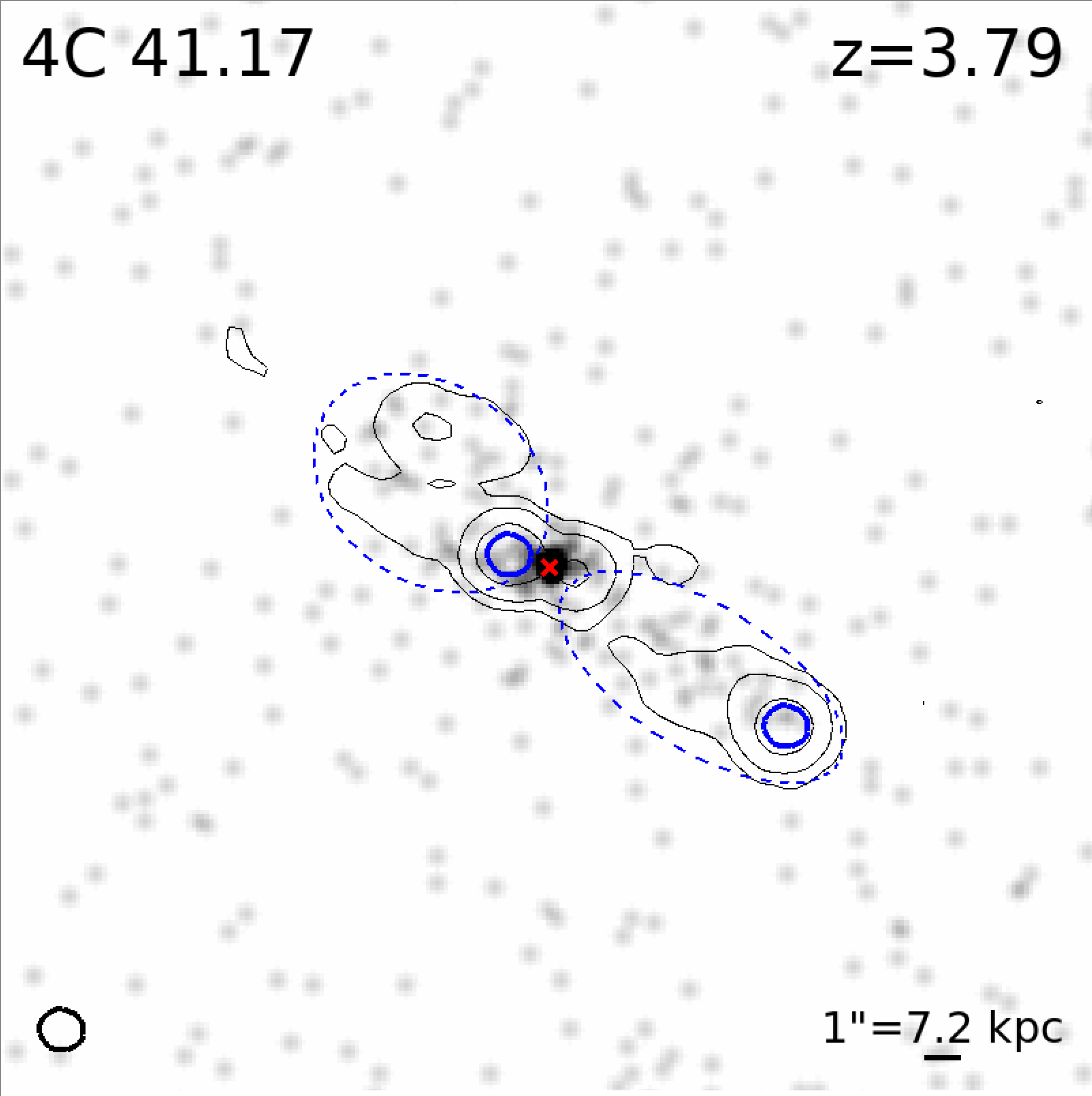}
    \includegraphics[height=2in]{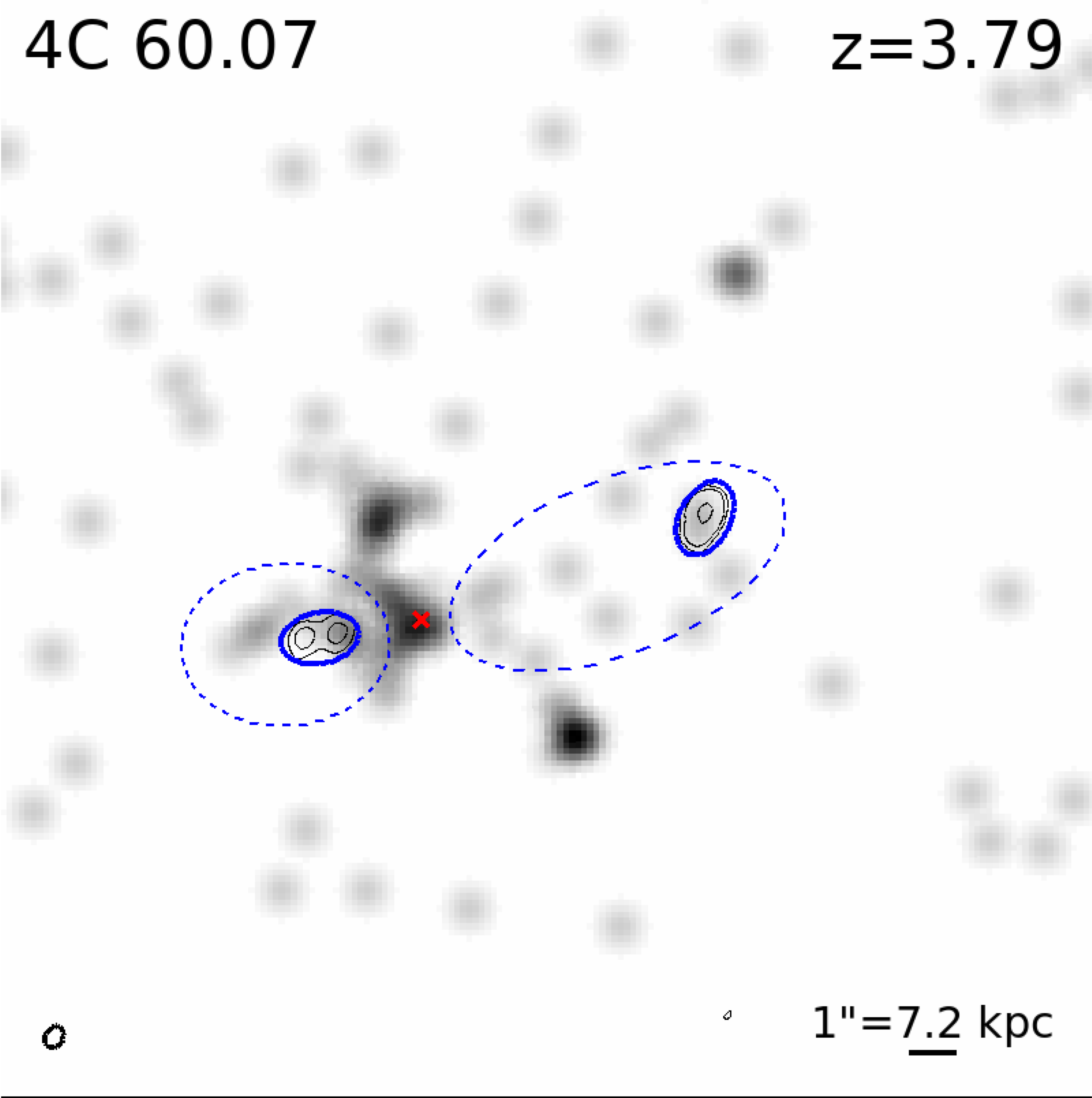}
    \includegraphics[height=2in]{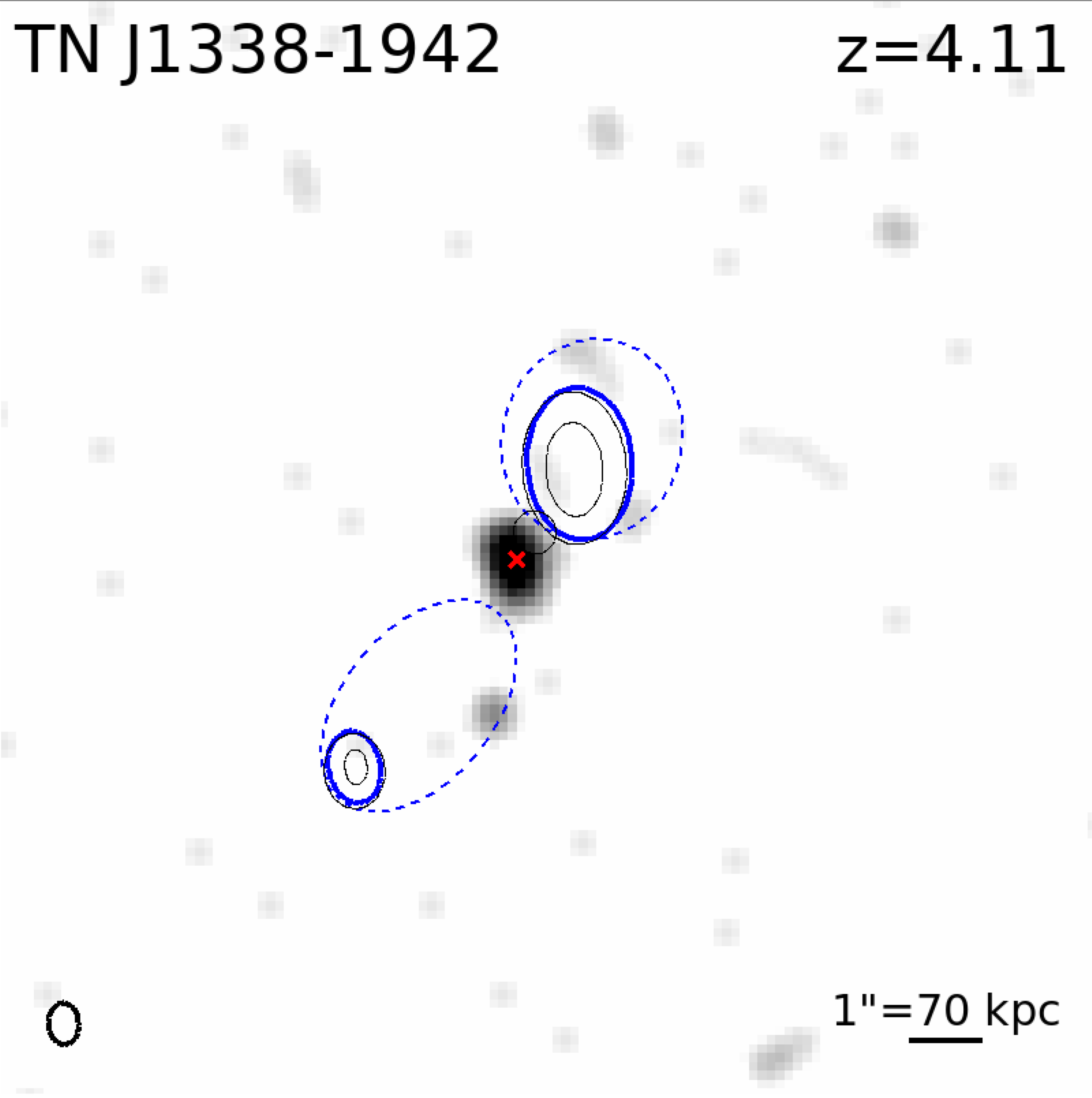} \includegraphics[height=2in]{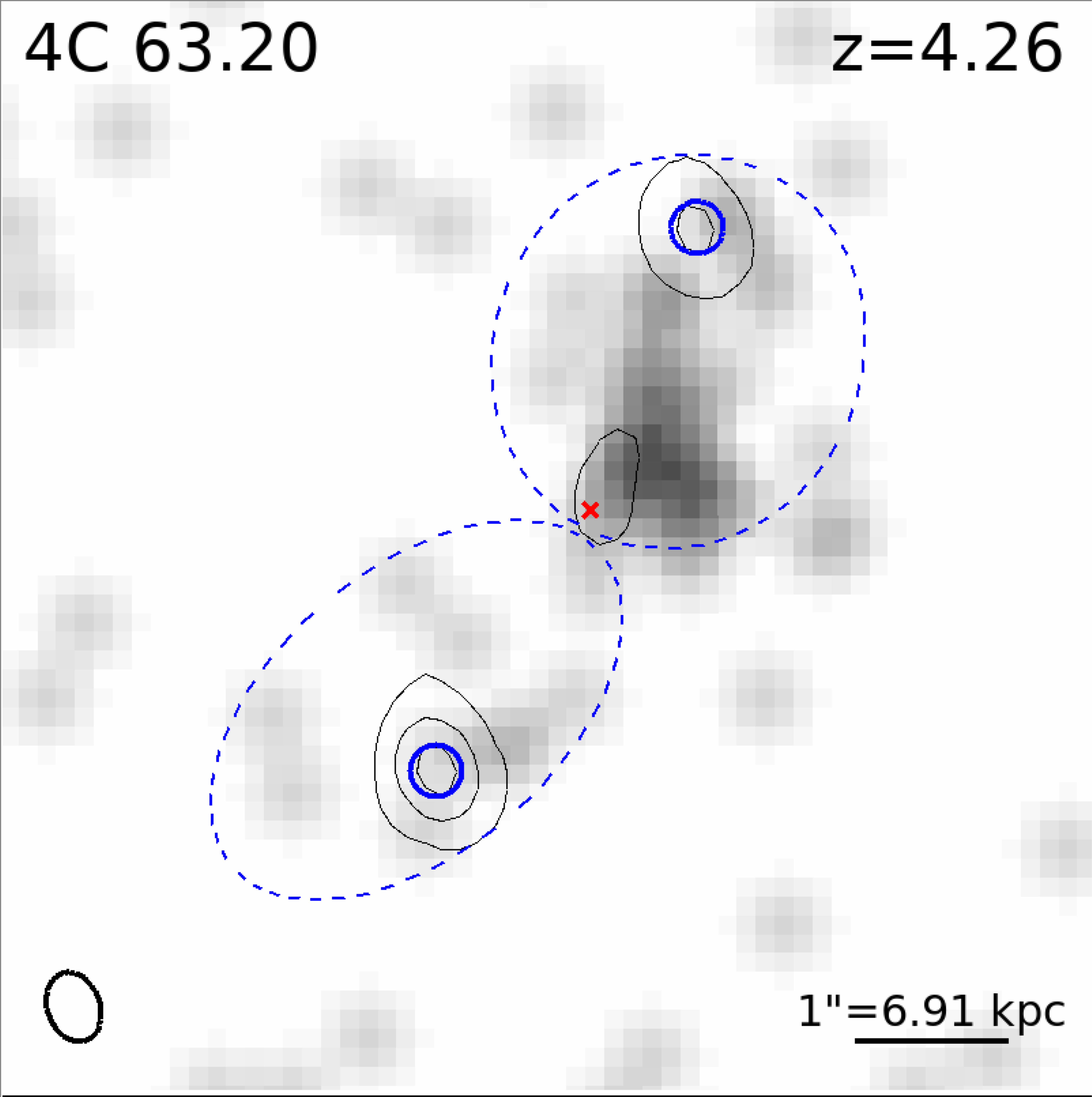}
    \caption{\label{fig:images}
    Images of 11 \hzrgs\ with public \cxo\ data. The grey scale image is the 0.5-8~keV \cxo\ image, smoothed with a Gaussian kernel to enhance low intensity emission. Radio contours are overlaid in black (the radio beam is depicted as a black ellipse in the left corner of each image), while the lobe and hot spot regions used in the text are shown as blue dashed and thick ellipses, respectively. We excise the core emission, and assume the total ``lobe'' volume when computing the total lobe$+$hot spot emission (see text).
    Radio contours are listed in units of mJy~per beam: 3C~469.1 (3, 18), 4C~$+$39.24 (1, 10), 3c~9 (4, 50, 200), 4C~$+$23.56 (0.8, 3), B3~0727$+$409 ($<$2, non-detection), 4C~$+$03.24 (1, 5, 10), 4C~$+$19.71 (4, 20), 4C~$+$41.17 (0.5, 1, 25), 4C~$+$60.07 (0.4, 0.8, 10), TN~J1338$-$1942 (1, 10), and 4C~$+$63.20 (1, 5, 25). 
    }
\end{figure*}
\section{Radio and X-ray luminosity measurements}
\label{section:data}
\subsection{High-redshift radio galaxy sample}
The working sample includes 11 \hzrgs\ observed with \cxo\ (an additional system, TN~J0924$-$2201, has a proprietary \cxo\ observation). We loosely define a \hzrg\ as a radio galaxy at $z>1$, and the goal is not to have a complete sample of \hzrgs\ (which does not exist in the \cxo\ archive), but rather to test whether the lobes are in equipartition, and thus brighten as $(1+z)^4$ due to IC/CMB. As such, the sample is biased towards \hzrgs\ luminous enough to be observed by \cxo, and we further reject systems where the X-ray emission cannot be reliably decomposed into a core and extended components, the system is embedded in a bright intra-cluster medium (at $z \lesssim 2$), or where the extended emission is only a jet. In these latter systems, it is likely that much or all of the X-ray emission is synchrotron emission from a very energetic population of cosmic rays \citep{meyer15,Georganopoulos2016}, which confounds the comparison we want to make \citep[however, see][for examples of IC/CMB jet emission]{Worrall2020}. Examples of rejected \hzrgs\ include 3C~294 ($z=1.779$), which is surrounded by bright thermal emission, and 3C~191 ($z=1.956$), whose diffuse radio emission occurs in a region severely contaminated by the image of the core (i.e., the point-spread function wings).   

Despite these limitations, the sample is sufficient to ask whether \hzrgs\ are similar to more local analogs and to estimate how close they are to equipartition. The \hzrgs\ in the sample, along with the radio and X-ray data we used, are listed in Table~\ref{table:sample}. Radio and X-ray images with the relevant regions labeled are shown in Figure~\ref{fig:images}, and the measurements are presented in Tables~\ref{table:total}, \ref{table:decomposed_lobes}, and \ref{table:decomposed_hs}. 

\subsection{Radio Data}
We measured radio fluxes from Karl G. Jansky Very Large Array (VLA) images with resolution of about an arcsecond. The available data span a range of frequencies (1.4--25~GHz) and quality, and often there are no images from compact arrays or single-dish telescopes that allow us to measure the total flux. Hence, some of the diffuse lobe emission may be resolved out. We retrieved pipeline-processed images from the VLA data archive where they exist, and followed the standard pipeline using the {\sc casapy} software \citep{McMullin2007} to create {\sc clean}ed images in the other cases, using the bandpass and gain calibrators appropriate for each observation. The data sets used are listed in Table~\ref{table:sample}.

\begin{figure*}
    \centering
    \hspace{-1cm}\includegraphics[trim=50 350 10 80,clip,width=0.75\textwidth]{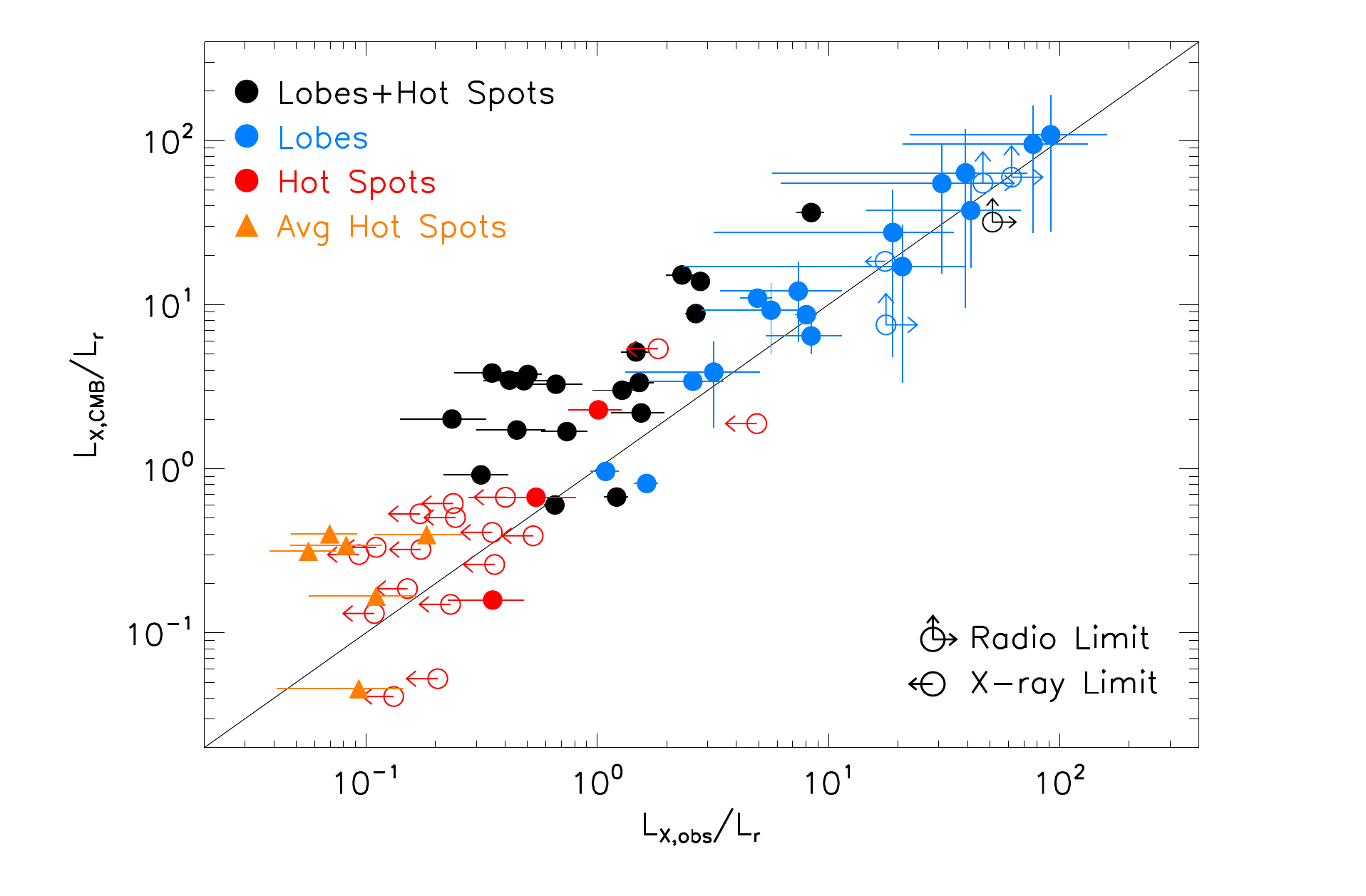}
    \caption{\label{fig:lx_lr}
    A comparison between the expected X-ray luminosity arising from IC/CMB in equipartition, \lxcmb, and the observed X-ray luminosity, \lxobs, each normalized by the measured radio luminosity, for the \hzrgs\ under analysis. For each galaxy, the black points (total) refer to the integrated emission (that is, the lobes plus hot spots, excluding the core), while the red and blue points are measurements from the  individual hot spots and lobes, respectively. Filled symbols represent sources detected in both the X-ray and radio bands, while open symbols refer to upper limits. Orange triangles represent mean hot spot values for systems where at least one hot spot is undetected. The upper right corner corresponds to large, weakly magnetized structures and the lower left corner to more compact, strongly magnetized structures. 
    Whereas the lobes are {scattered close to} the solid line that indicates agreement between the IC/CMB equipartition model predictions and the observations, the total emission is systematically offset from it. 
    }
\end{figure*}
\begin{table*}
\centering
\caption{Radio galaxies examined in this work. \label{table:sample}}
\begin{tabular*}{\textwidth}{@{\extracolsep{\fill}} lllc|clc|clc}
\hline
\hline
{Name} & {R.A.} & {Dec.} & {$z$} & {VLA Program} & {Date} & {Freq.} & {\cxo\ ObsID} & {Date} & {Sum GTI} \\
{} & {(J2000)} & {(J2000)} & & & & (GHz) & & & (ks) \\
\hline
3C~469.1	& 23:55:23.32	& $+$79:55:19.6	& 1.336	& AR0125    & 1985-02-21	& 1.49	& 9260 & 2009-05-24	& 20.2 \\
	&	&	&	& AR0125    & 1985-06-01	& 4.89\\			
4C~$+$39.24	& 09:08:16.919	& $+$39:43:26.3	& 1.883	& AB1093    & 2004-09-18   	& 1.45	& 5265    & 2005-03-02	& 19.9 \\
	& 	& 	& 	& AA0150    & 1993-04-18    & 1.45\\	
3C~9    & 00:20:25.219  & $+$15:40:54.59    & 2.02  & AL280     & 1992-12-13    & 1.55  & 1595    & 2001-06-10 & 82.6 \\
    &   &   &   &   &   &   & 17088 & 2015-11-02 & \\
    &   &   &   &   &   &   & 18700 & 2015-11-03 & \\
    &   &   &   &   &   &   & 18701 & 2015-11-04 & \\
4C~$+$23.56	& 21:07:14.82	& $+$23:31:45	& 2.483	& AC0374    & 1994-03-18	& 4.74	& 11687   & 2009-08-16	& 92.9 \\
	& 	& 	& 	& AR409 & 1999-05-28	& 4.84	\\		
B3~0727$+$409	& 07:30:51.346	& $+$40:49:50.8	& 2.5	& PERL2 & 1995-07-24	& 1.45	& 18184   & 2016-12-12	& 132.2 \\
	& 	& 	& 	& 	& 	& 	& 19959   & 2016-12-12	\\
	& 	& 	& 	& 	& 	& 	& 19960   & 2016-12-14	\\
4C~$+$03.24	& 12:45:38.364	& $+$03:23:20.7	& 3.57	& AM0336    & 1991-08-20	& 1.51	& 12288   & 2010-05-10	& 92.0 \\
	& 	& 	& 	& AC0374    & 1994-03-18	&  4.74	\\		
4C~$+$19.71	& 21:44:07.481	& $+$19:29:15.4	& 3.59	& AC0374    & 1994-03-18	& 4.74	& 12287   & 2010-08-23	& 91.7 \\
	& 	& 	& 	& AC0374    & 1994-03-18	& 8.24	& 13024 & 2010-08-26	\\
4c~$+$41.17	& 06:50:52.098	& $+$41:30:30.5	& 3.79	& AC0316    & 1992-12-16	& 1.59	& 3208    & 2002-09-25  & 149.3 \\
	& 	& 	& 	& AC0316    & 1992-12-16	& 4.74	& 4379  & 2002-09-26	& \\	
4C~$+$60.07	& 05:12:55.177	& $+$60:30:50.8	& 3.79	& AI88  & 2001-10-19	& 24.0	& 10489   & 2008-12-10	& 100.2\\
	& 	& 	& 	& AC0374    & 1994-03-18	& 4.74	\\		
TN~J1338$-$1942	& 12:38:26.1	& $-$19:42:31.1	& 4.11	& AD398 & 1997-01-25	& 4.86	& 5735    & 2005-08-29	& 78.7\\
	& 	& 	& 	& 	& 	& 	& 6367  & 2005-08-31	\\
	& 	& 	& 	& 	& 	& 	& 6368  & 2005-09-03	\\
4C~$+$63.20	& 14:36:37.326	& $+$63:19:13.1	& 4.26	& AC0374    & 1994-03-18	& 4.74	& 18106   & 2017-06-14  & 99.1 \\
	& 	& 	& 	& AB870 & 1998-08-17	& 4.89	& 19954 & 2017-03-13	\\	
	& 	& 	& 	& 	& 	& 	& 20033 & 2017-03-11	\\
\hline
\hline
\multicolumn{10}{p{0.9\textwidth}}{Notes. Positions and redshifts are from the NASA/IPAC Extragalactic Database (NED).}
\end{tabular*}
\end{table*}
%
\begin{table*}
\centering
\caption{X-ray and Radio Parameters for the integrated emission. \label{table:total}}
\begin{tabular*}{0.65\textwidth}{@{\extracolsep{\fill}} lclcccc}
\hline
\hline
{Name} & {Redshift} & {Position} & {$r_{\text{eff}}$} & {$\log L_{\text{r}}$} & {$\log L_{\text{X}}$} & {$\log$\lxcmb}  \\
&&& (kpc) & (erg s$^{-1}$) & (erg s$^{-1}$) & (erg s$^{-1}$)\\
\hline
3C~469.1	& 1.336	& N	& 67.2	& 44.09$\pm$0.02	& 43.71$\pm$0.07	& 44.63	\\
        	&	    & S	& 67.9	& 44.05$\pm$0.03	& 43.75$\pm$0.07	& 44.62	\\
4C~$+$39.24	& 1.883	& E	& 86.2	& 43.27$\pm$0.06	& 44.19$\pm$0.07	& 44.83	\\
	        &   	& W	& 80.2	& 43.84$\pm$0.02	& 44.21$\pm$0.06	& 45.02	\\
3C~9        & 2.02  & NE    & 18.9 & 44.45$\pm$0.01   & 43.63$\pm$0.06 & 44.29 \\  
            &       & SW    & 22.1 & 44.76$\pm$0.01   & 43.60$\pm$0.05 & 44.53 \\
4C~$+$23.56	& 2.483	& NW	& 36.7	& 43.81$\pm$0.04	& 44.24$\pm$0.05	& 44.75	\\
	        &   	& SE	& 47.4	& 43.80$\pm$0.02	& 44.25$\pm$0.04	& 44.94	\\
B3~0727$+$409	    & 2.5	& W	& 35.6	& $<$42.80	& 44.52$\pm$0.03 & 44.32	\\		
4C~$+$03.24	& 3.57	& NE	& 12.2	& 44.01$\pm$0.05	& 44.12$\pm$0.13	& 44.49	\\
	        &   	& S	& 11.5	& 44.38$\pm$0.04	& 44.25$\pm$0.11	& 44.60	\\
4C~$+$19.71	& 3.59	& N	& 12.1	& 44.33$\pm$0.01	& 43.70$\pm$0.20	& 44.63	\\
        	&	    & S	& 14.3	& 44.14$\pm$0.02	& 43.82$\pm$0.18	& 44.67	\\
4C~$+$41.17	& 3.79	& NE	& 19.5	& 44.36$\pm$0.01	& 44.53$\pm$0.06	& 45.07	\\
	        &   	& SW	& 13.7	& 44.23$\pm$0.01	& 44.41$\pm$0.07	& 44.75	\\
4C~$+$60.07	& 3.79	& E	& 11.8	& 44.05$\pm$0.01	& 43.88$\pm$0.16	& 44.57	\\
        	&   	& W	& 14.7	& 44.22$\pm$0.01	& 43.77$\pm$0.16	& 44.81	\\
TN~J1338$-$1942	& 4.11	& N	& 8.2	& 44.26$\pm$0.01	& 43.92$\pm$0.18	& 44.50	\\
        	&   	& S	& 7.8	& 42.96$\pm$0.03	& $<$43.64	& 43.90	\\
4C~$+$63.20	& 4.26	& N	& 7.7	& 44.09$\pm$0.01	& 44.28$\pm$0.13	& 44.43	\\
        	&	    & S	& 6.9	& 44.60$\pm$0.01	& 44.10$\pm$0.16	& 44.57	\\
\hline
\hline
\multicolumn{7}{p{0.6\textwidth}}{Notes. $r_{\text{eff}}$ is the radius of a sphere with the same volume as the emission region, assuming an axisymmetric, prolate spheroid. $L_{\rm r}$ and $L_{\rm X}$ refer to the measured radio and X-ray luminosity, respectively. $L_{\rm X,CMB}$ refers to the expected X-ray luminosity arising from IC/CMB (see Equation~\ref{eqn:beq}). }
\end{tabular*}
\end{table*}
%

\begin{table*}
\centering
\caption{X-ray and Radio Parameters for Lobes. \label{table:decomposed_lobes}}
\begin{tabular*}{0.75\textwidth}{@{\extracolsep{\fill}} lclcccccc}
\hline
\hline
{Name} & {Redshift} & {Side} & {$r_{\text{eff}}$} & {$\log L_{\text{r}}$} & {$\log L_{\text{X}}$} & {$\log$\lxcmb} & {$B_{\text{eq}}$} & {$\tau_{\text{sync}}$}  \\
{ } & { } & { } & {(kpc)} & {(erg s$^{-1}$)} & {(erg s$^{-1}$)} & {(erg s$^{-1}$)} & {($\mu$G)} & {(Myr)} \\
\hline
3C~469.1	& 1.336	& N	& 34.2	& 42.80$\pm$0.31	& 43.67$\pm$0.09	& 43.89	& 5.3 & 106 \\
	&	& S	& 30.2	& 42.91$\pm$0.27	& 43.66$\pm$0.09	& 43.87	& 6.1 & 86 \\
4C~$+$39.24	& 1.883	& E	& 74.05	& 42.34$\pm$0.54	& 44.23$\pm$0.07	& 44.32	& 2.9 & 263\\
	&	& W	& 74.1	& 42.25$\pm$0.59	& 44.21$\pm$0.08	& 44.28	& 2.7 & 293 \\
3C~9  & 2.02  & NE    & 18.9 & 44.32$\pm$0.01 & 44.53$\pm$0.06 & 44.23 & 34.4 & 6.4 \\
    &       & SW    & 22.1 & 44.40$\pm$0.01 & 44.43$\pm$0.07 & 44.38 & 31.6 & 7.3\\
4C~$+$23.56	& 2.483	& NW	& 36.7	& 42.94$\pm$0.76	& 44.22$\pm$0.05	& 44.38	& 7.9 & 59\\
	&	& SE	& 47.4	& 42.64$\pm$0.82	& 44.24$\pm$0.04	& 44.45	& 5.2 & 110 \\
B3~0727$+$409	& 2.5	& W	& 35.6	& $<$42.52	& 44.31$\pm$0.04	& 44.04	& $<$7.1 & $>$69 \\				
4C~$+$03.24	& 3.57	& NE	& 8.3	& $<$42.81	& 44.06$\pm$0.14	& 43.69	& 25.9 & 9.9\\
	&	& S	& 11.4	& 43.74$\pm$0.34	& 44.25$\pm$0.11	& 44.33	& 36.1 & 6.0\\
4C~$+$19.71	& 3.59	& N	& 12.1	& $<$42.03	& $<$43.58	& -	& 11.1 & 35\\
	&	& S	& 14.3	& $<$42.03	& 43.70$\pm$0.20	& 43.77	& 9.7 & 43\\
4C~$+$41.17	& 3.79	& NE	& 19.5	& 43.79$\pm$0.03	& 44.48$\pm$0.07	& 44.83	& 23.6  & 11\\
	&	& SW	& 13.7	& 43.51$\pm$0.04	& 44.41$\pm$0.08	& 44.44	& 26.5 & 9.5 \\
4C~$+$60.07	& 3.79	& E	& 11.8	& 42.20$\pm$0.35	& 43.82$\pm$0.18	& 43.78	& 12.8 & 28 \\
	&	& W	& 14.7	& 44.20$\pm$0.55	& 43.69$\pm$0.19	& 43.94	& 9.7 & 43\\
TN~J1338$-$1942	& 4.11	& N	& 8.2	& 42.52$\pm$0.71	& 43.84$\pm$0.20	& 43.75	& 21.5 & 13 \\
	&	& S	& 7.8	& 42.40$\pm$0.83	& $<$43.64	& 43.66	& 20.7 & 14 \\
4C~$+$63.20	& 4.26	& N	& 7.7	& 43.26$\pm$0.11	& 44.19$\pm$0.13	& 44.07	& 37.0 & 5.8\\
	&	& S	& 6.9	& 43.60$\pm$0.04	& 44.02$\pm$0.19	& 44.14	& 50.9 & 3.6\\
\hline
\hline
\multicolumn{8}{p{0.5\textwidth}}{Notes. Same notation as in Table 2 for \lx\ and \lr. $\tau_{\text{sync}}$ is calculated at 1~GHz.}
\end{tabular*}
\end{table*}

\begin{table*}
\centering
\caption{X-ray and Radio Parameters for Hot Spots. \label{table:decomposed_hs}}
\begin{tabular*}{0.75\textwidth}{@{\extracolsep{\fill}} lclcccccc}
\hline
\hline
{Name} & {Redshift} & {Side} & {$r_{\text{eff}}$} & {$\log L_{\text{r}}$} & {$\log L_{\text{X}}$} & {$\log$\lxcmb} & {$\log$\lxssc} & {$B_{\text{eq}}$} \\
{ } & { } & { } & {(kpc)} & {(erg s$^{-1}$)} & {(erg s$^{-1}$)} & {(erg s$^{-1}$)} & {(erg s$^{-1}$)} & {($\mu$G)}\\
\hline
3C~469.1	& 1.336	& N	& 15.9	& 44.07$\pm$0.02	& $<$43.04	& 43.55 & 42.84 & 33.9 \\
	&	& S	& 16.1	& 44.01$\pm$0.01	& $<$43.06	& 43.53 & 42.74 & 32.3 \\
4C~$+$39.24	& 1.883	& E	& 27.6	& 43.18$\pm$0.04	& $<$43.45	& 43.92 & 41.28 & 12.2 \\
	&	& W	& 27.4	& 43.83$\pm$0.01	& 43.83$\pm$0.13	& 44.19 & 42.29 & 18.7\\
3C~9  & 2.02  & NE    & 5.0 & 43.89$\pm$0.01 & $<$43.30 & 43.06 & 43.90 & 80.4\\
    &       & SW    & 8.4 & 44.50$\pm$0.01 & 43.61$\pm$0.20 & 43.71 & 43.71 & 78.0     \\
4C~$+$23.56	& 2.483	& NW	& 6.8	& 43.75$\pm$0.05	& $<$42.98	& 43.47 & 42.34 & 56.6\\
	&	& SE	& 3.7	& 43.77$\pm$0.02	& $<$42.95	& 43.04 & 42.56 & 95.8\\
B3~0727$+$409	& 2.5	& W	& 12.3	& $<$42.16	& 44.11$\pm$0.07	& 43.27 & 40.14 & $<$12.2 \\				
4C~$+$03.24	& 3.57	& NE	& 3.5	& 43.91$\pm$0.02	& $<$43.46	& 43.52 & 42.80 & 111.2 \\
	&	& S	& 2.9	& 43.72$\pm$0.02	& $<$43.45	& 43.31 & 42.56 & 113.9 \\
4C~$+$19.71	& 3.59	& N	& $<$1.2	& 44.32$\pm$0.01	& $<$43.44	& 42.94 & 43.77 & $>$354.6 \\
	&	& S	& $<$1.2	& 44.13$\pm$0.02	& $<$43.44	& 42.85 & 43.47 & $>$312.7 \\
4C~$+$41.17	& 3.79	& NE	& $<$3.8	& 43.79$\pm$0.29	& 43.53$\pm$0.01	& 43.62 & 42.82 & $>$95.4 \\
	&	& SW	& $<$3.8	& 43.79$\pm$0.01	& $<$43.39	& 43.62 & 42.82 & $>$95.4 \\
4C~$+$60.07	& 3.79	& E	& 3.9	& 44.05$\pm$0.01	& $<$43.44	& 43.75 & 43.23 & 109.8 \\
	&	& W	& 3.4	& 44.20$\pm$0.01	& $<$43.44	& 43.71 & 43.51 & 312.7 \\
TN~J1338$-$1942	& 4.11	& N	& 4.2	& 44.18$\pm$0.01	& $<$43.56	& 43.97 & 43.16 & 113.3 \\
	&	& S	& 2.9	& 42.82$\pm$0.03	& $<$43.51	& 43.10 & 41.15 & 65.0 \\
4C~$+$63.20	& 4.26	& N	& $<$2.2	& 44.07$\pm$0.01	& $<$43.63	& 43.49 & 43.20 & $>$184.9\\
	&	& S	& $<$2.2	& 44.59$\pm$0.01	& $<$43.63	& 43.71 & 44.02 & $>$260.0\\
\hline
\hline
\multicolumn{9}{p{0.5\textwidth}}{Notes. Same notation as in Table 2; $L_{\rm X,SSC}$ is the expected X-ray luminosity arising from synchrotron-self-Compton (based on the SSC emissivity expression given in Equation~\ref{eqn:ssc}).}
\end{tabular*}
\end{table*}

\begin{table*}
\centering
\caption{X-ray and Radio Parameters for Averaged Hot Spots. \label{table:avg_hs}}
\begin{tabular*}{0.61\textwidth}{@{\extracolsep{\fill}} lcccccc}
\hline
{Name} & {Redshift} & {$r_{\text{eff}}$} & {$\log L_{\text{r}}$} & {$\log L_{\text{X}}$} & {$\log$\lxcmb} & {$\log$\lxssc}  \\
&& (kpc) & (erg s$^{-1}$) & (erg s$^{-1}$) & (erg s$^{-1}$) \\
\hline
3C~469.1	& 1.336	& 16.0	& 44.05$\pm$0.01	& 42.80$\pm$0.56	& 43.54	& 42.54 \\

4C~$+$23.56	& 2.483	& 5.3	& 43.76$\pm$0.03	& 42.67$\pm$0.71	& 43.29 & 42.43 \\

4C~$+$03.24	& 3.57	& 3.8	& 43.83$\pm$0.01	& 43.09$\pm$0.86	& 43.43	& 42.88 \\

4C~$+$19.71	& 3.59	& $<$1.2	& 44.24$\pm$0.01	& 43.21$\pm$0.31	& 42.90	& 43.64 \\

4C~$+$60.07	& 3.79	& 3.7	& 44.13$\pm$0.01	& 42.97$\pm$0.55	& 43.73	& 43.13 \\

4C~$+$63.20	& 4.26	& $<$2.2	& 44.41$\pm$0.01	& 43.45$\pm$0.49	& 43.63 & 43.72 \\

\hline
\multicolumn{7}{p{0.58\textwidth}}{Notes. Average values are reported for systems where the hot spots are individually undetected in the X-rays, but for which adding up counts from hot spots on both sides of the core leads to a significant detection. The same notation is used as in Table 2 and 3.}
\end{tabular*}
\end{table*}
We measured fluxes from lobes and hot spots as described below, and report rest-frame luminosities as:  
\begin{equation}
    L_{\nu,\text{rest}} = 4\pi d_{\text{L}}^2 (1+z)^{\alpha-1} (\nu_{\text{rest}}/\nu_{\text{obs}})^{-\alpha} F_{\nu}
\end{equation}
where $\alpha$ is defined so that $F_{\nu} \propto \nu^{-\alpha}$ and is measured in the GHz band. We report luminosities as $\nu L_{\nu}$. Note that for $\alpha=1$, the redshift dependence cancels out. High-$z$ radio galaxies are sometimes ultra-steep-spectrum sources, so $\alpha$ may be larger in the observed-frame GHz band than for the electrons most relevant to IC/CMB. However, at low frequencies the angular resolution is worse and decomposing the system into lobes and hot spots is not generally possible, so we cannot directly test this possibility in each component.

\subsection{X-ray Data}

We measured X-ray fluxes from \cxo\ Advanced CCD Imaging Spectrometer (ACIS) images. We retrieved the data from the archive and processed it into analysis-ready level$=$2 files using the Chandra Interactive Analysis of Observations \citep[{\sc ciao} v4.12;][]{Fruscione2006} software. This involved using the {\sc ciao} \textit{chandra\_repro} script to reduce and calibrate the data, merging data sets for the same target where appropriate (i.e., same detector, exposure time, etc.), restricting the energy range from 0.5--8.0~keV to maximize the signal, and astrometrically registering the X-ray data against the radio images or optical catalogs. We then used the same regions defined before to measure X-ray count rates, from which we subtracted a mean on-field background. In the case of non-detection, the upper limit was measured using the expected number of background counts as the mean of a Poisson distribution, then determining the number of counts at which there is a 99.73\% likelihood that a counts cluster is a real source. 

To convert the count rate to X-ray luminosities, we use the appropriate \cxo\ response for each epoch, account for pile-up where necessary, subtract a local background, and assume that all the emission is IC/CMB with a spectral index of $\alpha=1$ ($\Gamma=2$). We further assume that there is only absorption from the Galactic column density \citep[computed from the HI4PI survey;][]{HI4PI2016}, and that a rest-frame 2--10~keV luminosity is appropriate for comparison to the radio data. In some \hzrgs\ {at $z \lesssim 2$} there is known to be X-ray emission from intra-cluster plasma. It is also possible that a source embedded in a denser medium will be surrounded by a cocoon of shock-heated gas. However, the signal is insufficient to extract a high fidelity spectrum, and in the cases where we can extract a spectrum it is consistent with a power law.

\subsection{Measurement Regions}

Most of the systems included in this study are double-lobed radio galaxies much larger than the \cxo\ point-spread function, so we make measurements on both sides of the core. We ignore or mask the core, and measure the total \lx\ and \lr\ on each side, as well as in the lobes and hot spots separately. The measured values are reported in Tables~\ref{table:total}, \ref{table:decomposed_lobes} and \ref{table:decomposed_hs}. The 1$\sigma$ error bars are purely statistical and do not account for error in redshift, region, size, spectral index, etc.

To determine the sizes and locations of the lobe and hot spot regions we use interferometric VLA radio images. If possible, we use the lowest frequency and multiple images with different resolutions to estimate the lobe extent, as the diffuse lobe emission could be resolved out in extended VLA arrays. In several cases, there is little lobe structure visible in any existing image (Figure~\ref{fig:images}) and so we supplement with the X-ray images themselves and define the outline of the lobe region based on the extent of the X-ray surface brightness. In the event that the lobes are not clearly defined in the radio or X-ray, we demarcate their extent by the position of the outermost hot spots, which are presumed to trace terminal jet shocks, and on the sides by the extent of radio or X-ray emission (see discussion of this choice in Section~\ref{section:results}). We exclusively use ellipses and assume that the volumes are prolate and axis-symmetric spheroids with effective radius $r_{\rm eff}$. This choice is necessarily subjective and does not reflect the complex volumes sometimes seen in more local radio galaxies. It is also possible that there are more extended lobe structures only visible at low frequencies due to spectral ageing. The impact of these assumptions is discussed in \S~\ref{sec:assumptions}. 

We followed a similar procedure to measure the sizes of the hot spots, which are usually very well defined in the radio. In some \hzrgs\ there are multiple hot spots, and here we restrict our analysis to the outermost hot spots (although we mask all hot spots when measuring the lobe emission). For these hot spots, if the source is resolved then we use an aperture based on the contour. Otherwise, we determine whether the hot spot is pointlike in the highest resolution image available by comparing the flux between images of different beamsizes. If little or no flux is resolved out at higher resolution, we use the smallest beamsize as the upper bound on the hot spot size and report the peak pixel flux in Jy (since the maps are calibrated in Jy~bm$^{-1}$). The hot spot regions are frequently comparable to or smaller than the ACIS half-power diameter of 0.8~arcsec. In this case, we treat the hot spot as an X-ray point source and measure the number of counts within a 2~arcsec diameter circle centered on the hot spot. We likewise use this value to estimate the X-ray flux from the lobes (i.e., subtract out any hot spot or point-like emission). However, to calculate the predicted X-ray luminosity we use the ``true'' volume from the radio. 

The reported X-ray luminosities are not corrected for the fraction of each lobe covered by the hot spots. If we assume uniform surface brightness, the correction factors are between 1-10\%, except in the northern lobe of TN~J1338$-$1942, where the correction factor is about 40\%. Not including this prediction may lead to systematically lower \lxobs\ relative to the predicted \lxcmb, but as the lobes may not have uniform surface brightness the true correction factor may be larger or smaller. Regardless, not adjusting \lxobs\ upwards does not substantially affect our results.

Because $B_{\text{eq}} \propto r^{12/7}$, the uncertainty in the volume is a large source of uncertainty in comparing the observed and predicted X-ray luminosities. It is also possible that the volume filling factor of the lobes is substantially less than unity, at least for relativistic cosmic rays, as low redshift, FR~II radio galaxies frequently show plasma flowing back from the terminal shocks into the lobes with jellyfish-like structures. If we have overestimated the volume, the effect is to over-predict the X-ray luminosity for a given radio luminosity. We return to this in Section~\ref{section:results}. 

\subsection{Comments on individual \hzrgs}
\subsubsection{3C~469.1,~4C~$+$39.24,~4C~$+$23.56,~4C~$+$60.07,~TN~J1338$-$1942} 
These are classical Fanaroff-Riley (FR) Type~II \citep{Fanaroff1974} radio galaxies with no visible jet, weak lobe emission, and very bright hot spots. We define the lobe region based on the extent of the hot spots and the apparent width of the X-ray emission associated with the galaxy, which leads to long, narrow lobes. 
\subsubsection{3C~9} This is a classical double radio galaxy with bright radio lobes that neatly match up with the enhanced X-ray emission. The hot spots are defined based on high frequency images, whereas the lobe extent is taken from a 1.55~GHz image. 
\subsubsection{B3~0727+$409$} \citet{Simionescu2016} discovered this \hzrg\ in the X-rays, as it has no diffuse radio counterpart (it does have a radio core and knot in the same direction as the extended X-rays). The X-ray structure is consistent with a jet and a diffuse lobe, so we define the lobe as the region outside of the inner jet. There is an X-ray bright region at the end of the lobe that we identify as a hot spot, but with no radio emission this classification is questionable. In the following analysis, we treat the region as a hot spot, but if we instead treat the entire system as a lobe it does not alter our conclusions.
\subsubsection{4C~$+$03.24} The best interferometric images of 4C~$+$03.24 show diffuse radio emission beyond the hot spots, but the size of any lobe cannot be reliably measured. Thus, we base the size of the lobe on the smoothed X-ray image, shown in Figure~\ref{fig:images}. This leads to some uncertainty on the X-ray flux and volume for the lobe beyond that reported in Table~\ref{table:decomposed_lobes}, but it is clear that the X-ray emission extends beyond the bright radio contours.  
\subsubsection{4C~$+$19.71} This \hzrg\ has bright hot spots and no significant lobe emission. The X-ray emission appears to be extended beyond the hot spots, and so we define the lobes based on the apparent X-ray enhancement. We note that \citet{wu17} found no statistically significant evidence for diffuse X-ray emission around the core using circular annuli, but along the axis joining the hot spots there is a significant detection. However, this is only true when summing the north and south lobe regions, and it is not clear whether the cluster of X-ray counts to the south is part of the radio galaxy. Of the radio galaxies our sample, the measurements are least secure for 4C~$+$19.71.
\subsubsection{4C~$+$41.17} There is bright X-ray emission on both sides of the core, but to the northeast the hot spot is very close to the core, with diffuse emission beyond. It is not clear whether the northeast hot spot is truly a terminal shock, but as the radio lobe is well defined and overlaps the X-ray enhancement well \citep[see also][]{wu17}, we define the lobe region based on the radio contours.

\subsubsection{4C~$+$63.20} This system was recently studied by \citet{Napier2020}, who determined that there is significant X-ray emission associated with the lobes. We define the extent of the X-ray lobes based on their work and their apparent width, but note that the radio emission allows for narrower lobes.
%
\begin{figure*}
    \centering
    \hspace{0.2cm}\includegraphics[trim=60 360 60 250,clip,width=\textwidth]{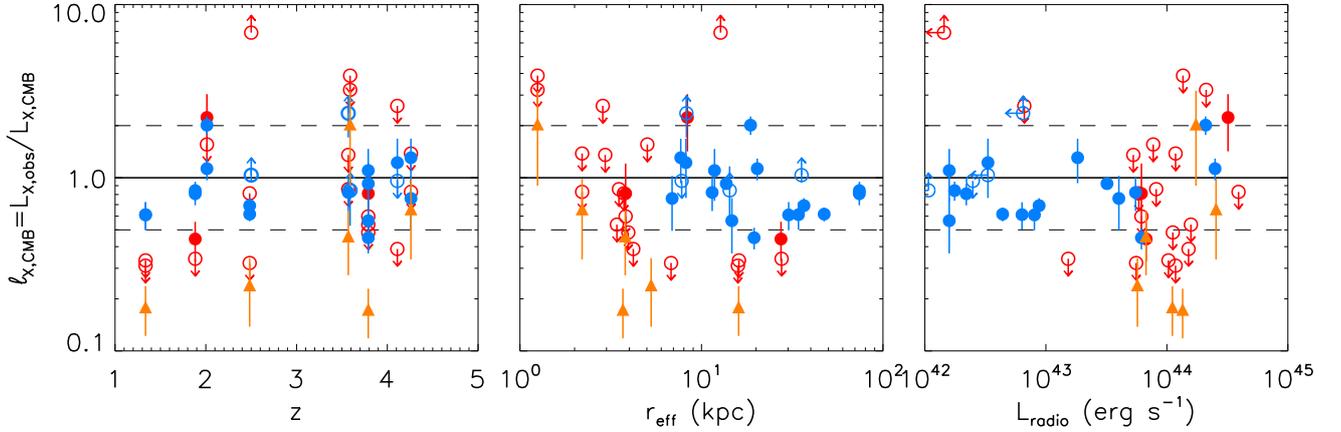}
    \caption{
    \label{fig:b_eq}
    The ratio \lxobs/\lxcmb\ for \hzrgs, plotted against redshift (\textit{left}), emitting region size (\text{center}), and measured radio luminosity (\textit{right}). Symbols are the same as in Figure~\ref{fig:lx_lr}, except that only the decomposed hot spots and lobes are shown. The measured luminosities of the lobes are within a factor of two (marked by the dashed lines) of the expected IC/CMB equipartition value. There is no significant correlation between the observed/predicted \lx\ and any parameter. The lobes of \hzrgs\ with a wide range of physical sizes and luminosities are near equipartition.}
\end{figure*}

\section{Results}
\label{section:results}
We began by computing the expected \lx\ from IC/CMB in equipartition, \lxcmb, following Equation~\ref{eqn:beq}, for each of the \hzrgs. The predicted total values, i.e. when the total (lobes$+$hot spot) volume and radio luminosity were used, are given in Table~\ref{table:total}, whereas the predicted values for the hot spots and lobes separately are given in Tables~\ref{table:decomposed_lobes} and \ref{table:decomposed_hs}. Most of the hot spots are not detected in the X-rays at the $3\sigma$ level, but in several systems there are counts in each of the hot spot regions, so we also compute an average value for those systems where stacking the hot spot regions across the core leads to a detection, using the same regions as defined above. The average measured luminosities, and predicted \lxcmb\ values for those are given in Table~\ref{table:avg_hs}. 

\subsection{Testing IC/CMB in equipartition}

Figure~\ref{fig:lx_lr} compares the observed X-ray emission, $L_{\rm X,obs}$, to the predicted X-ray luminosity arising from IC/CMB in equipartition, \lxcmb, each normalized by the measured radio luminosity. The black points (total) represent measurements from the entire radio galaxy (minus the core when present), while blue and red points represent the contribution from the individual lobes and hot spots, respectively.  Most lobes are detected in both X-rays and radio emission, albeit with some ambiguity as to what is true lobe emission (as opposed to a radiative jet or a faint hot spot complex). The southern lobe of TN~J1338$-$1942, which is not detected in X-rays, corresponds to an upper limit to the \lxobs/\lr\ ratio. Meanwhile, the lobe of B3~0727+409, the northeast lobe of 4C~$+$03.24, and the southern lobe of 4C~$+$19.71~are not detected in the radio, so we compute lower limits to the \lxobs/\lr\ and \lxcmb/\lr\ ratios. The northern lobe of 4C~$+$19.71 is not detected in X-rays, nor in radio, and is thus omitted from the plot. Only three hot spots have both radio and X-ray detections.

The observed (radio-normalized) X-ray emission from the lobes$+$hot spots is typically lower than predicted from IC/CMB, with a median observed-to-predicted ratio ($\ell_{\rm X,CMB}\equiv$ \lxobs/\lxcmb, see also Figure~\ref{fig:b_eq}) of 0.29; this result is fully consistent with previous investigations \citep[e.g.,][]{smail12,smail13}, and indicates that either the lobes, or the hot spots, or both, are off of (and specifically sub-) equipartition. 

Emission from the lobes alone is consistent with $\ell_{\rm X,CMB}=1$, with a median value of $0.84$. Incidentally, the median value for the lobes is similar to the average $\ell_{\rm X,CMB} \simeq 0.7$ measured at lower redshift for powerful radio galaxies \citep{Croston2005,kataoka05,Worrall2009}. 
Although most hot spots are not individually detected in X-rays, the distribution of the limits implies that $\ell_{\rm X,CMB}<1$ in most of them. When considering the averaged hot spot values (orange triangles) along with individually detected hot spots, we find a median $\ell_{\rm X,CMB}$ of $0.46$, indicating that the hot spots are off of equipartition. However, as we discuss in the following subsection, it is not clear that we can conclude this because SSC is expected to contribute significantly to--possibly dominate--the X-ray emission in hot spots, and there may also be a second synchrotron population contributing to the X-rays. It is also possible that a failure to measure equipartition is due to offsets between the X-ray and radio peaks for a given hot spot, which can be significant fraction of an arcsecond and likely result from relativistic shocks \citep{hardcastle04,Hardcastle2007}\\

In light of this analysis, we argue that the lobes of the \hzrgs\ are fully consistent with the expectations from a IC/CMB model in equipartition. Previous claims to the contrary are typically based on luminosity measurements of the aggregated, lobe$+$hot spot regions, and such measurements are not usually appropriate, as interpreting the aggregate relies on a precise understanding of each component. Instead, a fair comparison between theory and observations requires substituting the relevant (i.e. lobe vs. hot spot) luminosities as well as sizes in Equation~\ref{eqn:beq}. This can only be achieved for high-quality data which allow for a proper decomposition. \\

Figure~\ref{fig:b_eq} also illustrates that there is no significant correlation (at the 95\% confidence level) between $\ell_{\rm X,CMB}$ and the redshift, effective radius, or radio luminosity of the \hzrgs\ under analysis. This is true of either the hot spots or the lobes, when considering limits as well as detected systems. 
The test is warranted, since, given a sufficiently large sample that includes very large systems  ($r_{\text{eff}} > 100$~kpc, i.e., in the realm of giant radio galaxies), we would expect to see an anti-correlation between $\ell_{\rm X,CMB}$ and $r_{\text{eff}}$. This is because the predicted scaling of $L_{\text{X,CMB}}$ with $r_{\text{eff}}^{12/7}$ assumes a uniform filling factor and magnetization. These assumptions are unlikely to hold for very large systems even if the \textit{emitting volume} is roughly in equipartition. However, we do not see this effect in the sample.\\

To summarize, Figures \ref{fig:lx_lr} and \ref{fig:b_eq} show that (i) the lobes of the \hzrgs\ considered in this work are consistent with IC/CMB at or near equipartition, and (ii) there is no indication that this changes with redshift, AGN power, or lobe size.  

\subsection{The role of Synchrotron Self-Compton}

SSC emission, whereby the relativistic electrons also cool by IC scattering off of the same population of photons that they produce via synchtotron, is expected to be important in the compact, highly magnetized hot spots. Thus, it is not surprising that $\ell_{\rm X,CMB}<1$ for most hot spots in our sample, since the X-rays are likely not predominantly IC/CMB.

Since the SSC X-ray luminosity, \lxssc, also depends on $B$, we can test whether $B \approx B_{\text{eq}}$ in the hot spots with this definition.
To calculate the equipartition \lxssc, we need to know both the number of seed photons (estimated from the radio luminosity) and the optical depth to Compton scattering, $\tau = \sigma_{\text{T}} N_0 r$, where $N_0$ is the number density of electrons, and $r$ is the characteristic path length, which we take to be the radius of a uniform sphere. In equipartition, $u_B = (1+k)u_e$, where $k = 0$ is the ratio of energy in non-radiating particles to electrons, $u_B = B_{\text{eq}}^2/8\pi$, and 
\begin{equation}
    u_e = \int E N(E) dE = \int E (N_0 E^{-p}) dE = N_0 m_e c^2 \int_{\gamma_{\text{min}}}^{\gamma_{\text{max}}} \gamma^{1-p} d\gamma
\end{equation}
where we have assumed a power law distribution of electron energies between $\gamma_{\text{min}} = 10$ and $\gamma_{\text{max}} = 10^4$, with index $p = 2$ (i.e., exactly the same assumptions as for IC/CMB). Thus, we can solve for $N_0$ in terms of \beq. 

The synchrotron emissivity is a power law between the rise and break regions, with $J_{\text{s}}(\nu) = J_{0,\text{s}} \nu^{-\alpha}$, where $\alpha = (p-1)/2$. The emissivity has units erg~s$^{-1}$~cm$^{-3}$~Hz$^{-1}$~sr$^{-1}$, so we can find $J_{0,\text{s}}$ from the measured \lr\ at a given frequency and the hot spot size, which is converted to an effective spherical radius. The SSC emissivity can then be written \citep{Ghisellini2013b}: 
\begin{equation}
\label{eqn:ssc}
    J_{\text{ssc}}(\nu) = \biggl(\frac{4}{3}\biggr)^{\alpha-1}\frac{1}{2} \tau J_{0,\text{s}} \nu^{-\alpha} \int_{\nu_{\text{min}}}^{\nu_{\text{max}}} \nu^{-1}d\nu
\end{equation}
We convert this back to a 2-10~keV X-ray luminosity for comparison with the data (Table~\ref{table:decomposed_hs}). 

\begin{figure}
    \centering
    \hspace{-0.5cm}\includegraphics[trim=50 350 50 80,clip,width=0.5\textwidth]{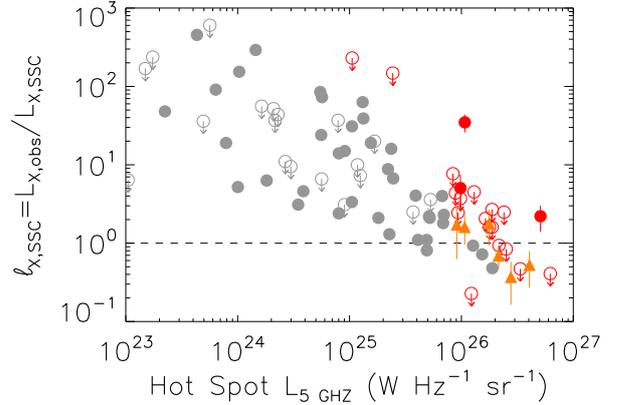}
    \caption{
    \label{fig:hardcastle}
    A comparison between the expected X-ray luminosity arising from SSC in equipartition, \lxssc, and the observed X-ray luminosity, \lxobs, for a sample of radio galaxy hot spots. 
    The grey points are from the low-$z$ sample of \citet{hardcastle04}, while the red points are the \hzrgs\ in this work. As in previous figures, the orange triangles represent the average values for pairs of opposing hot spots for systems in which the individual hot spots are not X-ray detected. The expected SSC luminosity was predicted following \citet{hardcastle04}, using the measured radio luminosity and emitting region size and assuming equipartition. The \hzrgs\ in our sample are broadly consistent with the \citet{hardcastle04} sample and with being close to equipartition.}
\end{figure}

When comparing the observed X-ray luminosity to \lxssc\ for the hot spots in our sample (Figure~\ref{fig:hardcastle}), the individual detections and limits (red circles) are broadly scattered around the expected equipartition value $\ell_{\rm X,SSC}\equiv$\lxobs/\lxssc$\simeq 1$, and the average hot spot measurements (orange triangles) are consistent with equipartition. The hot spots are also consistent with the trend found by \citet{hardcastle04} for low-$z$ systems that the discrepancy between the observed \lxobs\ and equipartition \lxssc\ is a function of radio luminosity.
Figure~\ref{fig:hardcastle} shows their measurements in grey and ours in red/orange. The hot spots from the \hzrgs\ are at the high radio luminosity end (at least in part due to selection effects), but they follow the same basic trend as the lower $z$ sample. This is particularly clear for the average hot spot values (i.e., those systems where we detected the hot spots in X-rays when stacking the two regions on either side of the core), which are shown as orange triangles. \\

To summarize, our analysis shows that both the lobes and hot spots of \hzrgs\ appear to be consistent with equipartition, and the hot spots (where the contribution from SSC is important in X-rays) are similar to the most luminous systems studied at lower redshift. In other words, the \hzrgs\ under examination are not unusual in their behaviour as radio galaxies. 

\subsection{Impact of assumptions}
\label{sec:assumptions}

We make several assumptions when measuring \lr, $r_{\text{eff}}$, and \lxobs. First, we assume that the lobes are ellipsoids with a filling factor of unity, and that the volume can be measured from the radio or X-ray surface brightness contours. A similar assumption applies to the hot spots. Secondly, we assume that there is negligible dimming of \lr, such that it is useful to calculate \beq. Thirdly, we assume the same spectral index $\alpha = 1$ in all systems. Lastly, our estimate of \beq\ assumes that the observed $\nu L_{\nu}$ represents all of the particles contributing to the energy density. We address the impact of (the uncertainty in) these assumptions on our results in turn below.

\lxcmb\ depends on $r_{\text{eff}}^{12/7}$, but the measured \lr\ and \lxobs\ also depend on the projected area of the lobe, so the effects largely cancel out if the lobe emission is approximately uniform. Despite the low numbers of counts in the lobes, most of the X-ray lobes have close to uniform surface brightness. Since the lobe regions are defined based on the surface brightness of the radio or X-ray emission, it is not likely that the regions we adopted are too small. However, they may be too large, with ellipses whose $r_{\text{eff}}$ differ by 25-30\% consistent with the data in several cases. This would lead to a 15\% error in $\ell_{\rm X,CMB}$. A similar uncertainty is incurred if the lobes are oblate, rather than prolate spheroids, but in the opposite direction. This amount of uncertainty does not change our conclusions. 

The other uncertainty in the volume is the filling factor, $\phi$, but as this uncertainty is related to the particle composition of the jet (i.e., whether there is a significant hadronic component), it pertains to the entire measurement scheme and should not be considered in isolation. \\

One of the objectives of this work is to examine whether IC/CMB cooling can dim radio lobes at high redshift, in which case the measured \lr\ can be smaller than what would be expected for a given \beq. Since we derive \beq\ from \lr, this would erroneously decrease \lxcmb\ by a factor of \lr$^{2/7}$. Since $\ell_{\rm X,CMB}$ explicitly contains the $(1+z)^4$ effect and $\ell_{\rm X,CMB}$ is not (further) correlated with $z$ (Figure~\ref{fig:b_eq}), we can conclude that any dimming must not be severe, with the caveat that it would be most prominent for the few systems at the largest $z$. Dimming consistent with the lack of correlation (by up to a factor of two at high $z$) would reduce $\ell_{\rm X,CMB}$ to $\approx$75\% of its nominal value, so this would not qualitatively change our conclusions. 

Thirdly, we have assumed that $\alpha=1$ in each source for calculating \lr. This is a reasonable approximation for many steep-spectrum radio sources, but there may be significant dispersion. Since the radio sources are unresolved below 1~GHz and there are generally too few X-ray counts to robustly constrain $\alpha$ from the X-rays, we cannot directly test whether \lr$\simeq \nu L_{\nu}$ in the GHz band is a good approximation for each source. However, based on the unresolved radio SEDs of sources with weak radio cores\footnote{available via NED}, only a few sources show any evidence for a synchrotron break below 1~GHz, in which case $\alpha \approx 1$ is a good approximation.

We can estimate the impact of not knowing $\alpha$ in each source. Since the systems are \hzrgs\ and we focus on extended components, we can place a lower bound of $\alpha = 0.7$. Meanwhile, no source appears to have $\alpha \gtrsim 2$, so we expect that $0.7 < \alpha < 2$, where these are conservative limits. If $\alpha = 0.7$ instead of 1, the rest-frame \lr\ is 40\% of the reported value, which in turn leads to a factor of 1.5 higher (true) $\ell_{X}$. If $\alpha = 2$ instead of 1, the true $\ell_{X}$ is 0.7 times the reported value. We do not expect the \hzrgs\ here to be predominantly towards one end or the other of this range, and so uncertainty in $\alpha$ will manifest as an overall uncertainty in \lxcmb\ of about 40\%. This would not change our conclusions, and indeed if treated as an ``error bar'' that can be added in quadrature with the statistical uncertainty, would have a small impact on the appearance of Figure~\ref{fig:lxlr_z}; however, since this uncertainty cannot be treated in this way, it is not included in that figure.

Lastly, we have assumed in calculating \beq\ that \lr\ represents the extent of the particles in each system, i.e., that $\gamma_{\text{min}}$ and $\gamma_{\text{max}}$ are based on the radio emission. This is not the case, so it is likely that \beq\ is underestimated, especially for steep-spectrum sources (although we do not know the true $\gamma_{\text{min}}$ and $\gamma_{\text{max}}$). We use \beq\ to estimate the predicted \lxcmb, but in order to compare our derived \beq\ values with those in the literature, Tables~\ref{table:decomposed_lobes} and \ref{table:decomposed_hs} include them. In the lobes, the values range from 2--50~$\mu$G, which is similar to the field strengths measured in more nearby samples using different techniques \citep{Croston2005,Birzan2008,Harwood2016}, and where similar radio--X-ray analyses have been carried out \citep[e.g.,][]{Croston2005}. This suggests that our assumption does not significantly affect our conclusions.

Another way to consider the same issue is to calculate the $\gamma$ probed at 1~GHz from \beq, using $\gamma \approx \sqrt{\nu/\nu_{\rm G}}$. The electron gyrofrequency is $\nu_{\rm G} \approx 2.8 B[\mu{\rm G}]$~Hz. 2,000 and 11,000, with an average near 4,000. This is near the upper end of the range frequently used in the literature, $10 < \gamma < 10^4$, so for steep-spectrum sources \beq\ may indeed be underestimated. If we adopt an initial electron energy power-law index of 2, we expect \beq\ to be higher by 10-40\% in the different lobes than our calculated values, based on the $\gamma$ that produces 1~GHz emission. Since we do not know the true range nor the electron energy distribution, it is not clear whether or how much this estimate is better, but it does not substantially alter our conclusions.

A related issue is that the X-ray bright sample observed by \cxo\ is likely biased towards very active sources, so that CMB cooling may not have had enough time to quench the system (i.e., the replacement rate of energetic particles is very high). We calculated the synchrotron lifetimes based on our \beq\ values (see Table~\ref{table:decomposed_lobes} for the lobes, calculated at 1~GHz):
\begin{equation}
    \tau_{\rm sync} = \frac{m_e c^2}{\tfrac{4}{3}\sigma_{\rm T} c U_{\rm B} \gamma}
\end{equation}
The values range from 3-300~Myr, reflecting the strong dependence on the lobe size (on which our inferred \beq\ depends). There is no significant correlation between the observed \lx\ or \lr\ and the projected lobe size, nor between the core luminosities and \beq, so there is no clear preference for especially active sources in the sample. Although the sample presented here is small and biased towards X-ray bright sources, the similarity to low-redshift radio galaxies in terms of size and \beq\ indicate that it is a reasonable sample in which to test the CMB-quenching hypothesis.

\section{Discussion}

\subsection{IC/CMB X-ray brightening}

\begin{figure*}
    \centering
    \hspace{-0.5cm}\includegraphics[trim=50 340 50 160,clip,width=\textwidth]{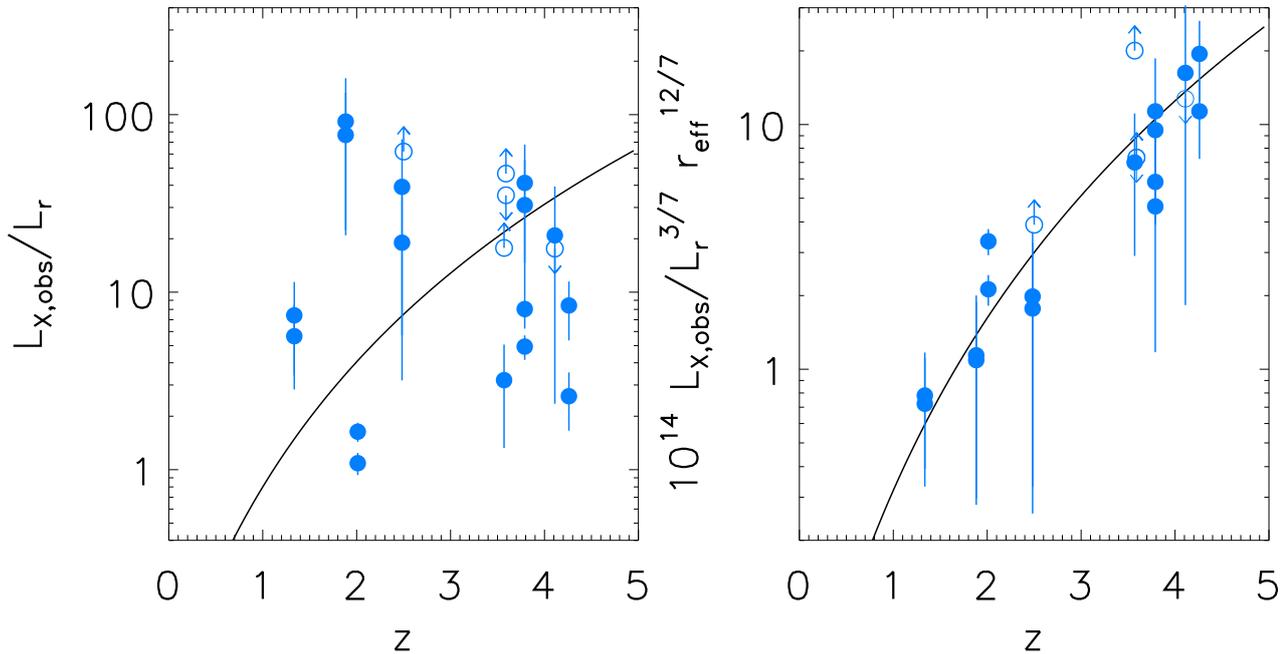}
    \caption{
    \label{fig:lxlr_z}
    Even in equipartition, we do not expect \lxobslr\ to correlate strongly with $z$ in small samples, because \lx\ also depends on $B^2$, which depends on the lobe size (see Equation \ref{eqn:beq}). 
    \textit{Left}: \lxobslr\ as a function of $z$ for the resolved lobes of the target \hzrgs. Open circles are upper limits. \textit{Right}: Scaled $L_{\rm X,obs}/(L_{\rm r}^{3/7} r_{\rm eff}^{12/7})$ as a function of $z$, for the same lobes. This is a re-projection of Figure~\ref{fig:lx_lr}, as the exponents on \lr\ and $r_{\rm eff}$ are those in Equation~\ref{eqn:beq}. In both panels, the solid line is proportional to $(1+z)^4$ and arbitrarily scaled. It is not a fit, but clearly the points in the right panel are more consistent with this scaling, and they are at least strongly correlated with $z$.
    }
\end{figure*}

Motivated by the mandatory increase in the expected IC/CMB X-ray luminosity at higher redshift, as $u_{\rm CMB}\propto (1+z)^4$, previous works have searched for a correlation between \lxobslr\ and $z$ (Section~\ref{section:intro}). However, here we wish to emphasize that, if \hzrgs\ are in equipartition, then one does not expect \lxobslr\ to strongly correlate with $z$ even though CMB X-ray brightening does occur. 
This is for two reasons: first, as we showed, any such correlation can be washed out when comparing entire radio galaxies, for which (at high redshift) most X-rays come from the lobes and most radio emission comes from the strongly magnetized jets or hot spots. Secondly, even in lobe-dominated sources, and/or if the emission from the hot spots is thoroughly excluded, \lxcmb\ depends also on \lr\ and size, as \lr$^{3/7}$ and $r^{12/7}$, respectively. Figure~\ref{fig:lxlr_z} illustrates this point by plotting \lxobslr\ for only the lobes in our sample, in the left panel, and a scaled version of Equation~\ref{eqn:beq}, in the right panel. The lines are arbitrarily scaled functions of $(1+z)^4$ and are not fits to the data. One can clearly see that, even within the lobes, \lxobslr\ is not strongly correlated with redshift, whereas the function in the right panel shows remarkable agreement with the expectations from IC/CMB in equipartition, where the measured $(L_{\rm X,obs}/L_{\rm r}^{3/7}r_{\rm eff}^{12/7})$ values for the lobes are consistent with a $(1+z)^4$ scaling up to $z\simgt 4$. 

In essence, this proves that, when decomposed into lobes and hot spots, the \hzrg\ radio and X-ray data are fully consistent with a picture where synchrotron cooling in the lobes is progressively offset by IC scattering of CMB photons, which in turn causes the lobes to brighten in the X-rays, and become dimmer in the radio band. This also implies that there is no need for an additional FIR photon field to explain the observed \lxobslr\ in the lobes of this sample, which overlaps the sample from \citet{smail13} where this solution was explored (even though we cannot rule out that IC/FIR does contribute a fraction of the X-rays from these data). 

An additional reason that surveys of \hzrgs\ may not find a correlation between \lxobslr\ and $z$ is that the X-ray emission in a significant fraction of these systems with \cxo\ data is dominated by jets (which were not included in our sample). There is evidence for IC/CMB dominated jets \citep{Worrall2020,Schwartz2020}, as well as X-ray synchrotron emission \citep{meyer15}. As with SSC emission in the hot spots, a second, more energetic population of electrons that radiates X-ray synchrotron emission will not depend on redshift. Hence, CMB brightening may not be detectable from \lxobslr\ even with a much larger sample of resolved \hzrgs\ than currently exists.

\subsection{IC/CMB radio quenching?}

The $(1+z)^4$ dependence of the IC/CMB effect has been invoked to explain the apparent dearth of luminous radio-loud AGN at $z\simgt 3$ \citep{volonteri11}. This work started by estimating the fraction of radio-loud objects amongst all FIRST\footnote{Faint Images of the Radio Sky at Twenty-cm.}-detected SDSS\footnote{Sloan Digital Sky Survey.} quasars (defined as AGN with bolometric luminosities exceeding $10^{47}$~erg~s$^{-1}$ in the seventh data release), in different redshift bins. This fraction was then (i) normalized to the radio-quiet quasar luminosity function derived by \cite{hopkins07}, to obtain the total number of radio-loud quasars per redshift bin between $1\simlt z \simlt 6$, and (ii) compared to the expected number of massive radio galaxies (with estimated black hole masses in excess of $10^9$ solar) starting from the parent blazar population. The latter was derived by normalizing the number of blazars inferred from those detected in the three-year Swift-BAT\footnote{Burst Alert Telescope.} sample \citep{ajello09} to the ``minimal" blazar luminosity function derived by \cite{ghisellini10}, then multiplied by a factor $2\Gamma^2$, where $\Gamma$ is the bulk Lorentz factor of the jet, to account for the misaligned sources. This analysis found agreement between the expected and observed numbers up to $z\simeq3$, above which the number of observed systems appears to drop sharply, with as many $500-5200$($5200-7800$) sources expected in the $5-6$($4-5$) redshift bin, vs. only 56(252) identified/detected, assuming $\Gamma=15$. 

Taking these numbers at face value, \citet{ghis_cel_tav14} and \cite{Ghisellini2015} explored the possibility of IC/CMB radio quenching, whereby IC/CMB becomes more effective than synchrotron cooling at high-$z$, thus causing radio dimming (at rest-frame GHz frequencies) to below the 1~mJy sensitivity of the FIRST survey. They did so by fitting the spectral energy distributions (SEDs) of a sample of blazars at $z>4$ with components for the accretion disk, beamed jet, and torus, then assessing the amount of lobe and hot spot emission allowed under equipartition. They concluded that IC/CMB quenching is possible when the kinetic luminosities of the hot spots and lobes are 1\% that in the jet, which is consistent with the radio data for their sample. In this case, hot spot emission is too weak (at 1.4~GHz) to be detected, and the lobes can be quenched by IC/CMB. 

Although the \hzrgs\ in our sample are at systematically lower redshifts, they are very similar to the misaligned counterparts expected for the high-$z$ blazars discussed in \citet{ajello09} and \citet{Ghisellini2015}. Their bolometric luminosities are between $10^{46}$ and $10^{48}$~erg~s$^{-1}$ (assuming a 2--10~keV bolometric correction factor of 50 for the core; \citealt{Vasudevan2007}), which places them in the same regime as the high-$z$ blazars studied by \citet{Ghisellini2015}. In this work, we also find that they are close to equipartition and that they appear to be very similar to low-$z$ radio galaxies. Meanwhile, \citet{Ghisellini2015} concluded that $z>4$ blazars are consistent with equipartition and that they are similar to lower redshift blazars. Taken together, these findings indicate that the \hzrgs\ presented here are indeed misaligned blazars in the same regime as those in \citet{Ghisellini2015}. 

Nevertheless, most of the \hzrgs\ in our sample have hot spots with radio luminosities exceeding $10^{43}$~erg~s$^{-1}$; as shown below, these cannot be quenched at $z<5$ (this was already discussed for a few of the systems in our sample, by \citealt{wu17} and \citealt{Napier2020}). For a hot spot with radius 2~kpc, $10^{43}$~erg~s$^{-1}$ corresponds to $B_{\text{eq}} \simeq 100$~$\mu$G, whose magnetic energy density is in equipartition with the CMB at $z=5$. Thus, no significant dimming will occur until higher redshifts. Meanwhile, the FIRST sensitivity of $\simeq$1~mJy at 1.4~GHz corresponds to $4\times 10^{42}$~erg~s$^{-1}$ at $z=5$, so this hypothetical hot spot would be detected by FIRST. Almost all of the hot spots in our sample have higher \lr\ values, so they would be detectable as FIRST sources if they were observed at $z\simgt 5$. 
This suggests that, even if IC/CMB quenching of lobes occurs, to the extent that our sample is representative of \hzrgs\ then IC/CMB quenching alone cannot explain the deficit of luminous, radio-loud quasars above $z\simgt 3$.

The main caveat with this statement is that the sample may not be representative: it consists of \hzrgs\ observed with \cxo, and for which it is straightforward to decompose the lobes and hot spots (generally implying that, by construction, the lobes are visible, and not completely quenched). However, we note that we use VLA images with angular resolution $\lesssim 1^{\prime\prime}$ (as compared to 5$^{\prime\prime}$ for FIRST), and the rms noise is frequently less than the 0.15~mJy of FIRST, so the decomposition presented here could not be achieved with the FIRST data. 
Regardless, the question is whether the \textit{hot spots} in our sample are unusually luminous. For a comparison, we consider the catalog of low-$z$, FR~II radio galaxies from the FIRST survey \citep{Capetti2017}. Most of those radio galaxies have well-defined hot spots that contribute a significant fraction of the GHz radio luminosity, implying that higher-redshift, jetted AGN are also likely to possess hot spots. For AGN with $L_{\text{bol}} \gtrsim 10^{47}$~erg~s$^{-1}$, these hot spots should typically be very luminous. We also note that \citet{Ghisellini2015} do not robustly constrain \lr\ at rest-frame 1~GHz in the hot spots, and most of the blazars in their sample do allow for hot spots that could be detected by FIRST while remaining consistent with the fits to the SED of the AGN and beamed jet. 

From this we are drawn to a second conclusion; whereas our work is consistent with IC/CMB quenching of most radio lobes at $z \simgt 4$, this effect would not be sufficient to explain the deficit of high-$z$, radio-loud AGN if such AGN frequently have bright hot spots, as suggested by our sample and a large number of low-$z$ radio galaxies detected in the FIRST survey. \\

If IC/CMB is not responsible for making those radio galaxies ``disappear", what then? Starting with the initial claim by \cite{volonteri11}, several scenarios have been proposed--and largely dismissed--including a drop of the average jet bulk Lorentz factor at high-$z$, and the possibility that the SDSS misses a large fraction of high-$z$, radio-loud AGN because of extreme obscuration. Here, we briefly consider additional possible explanations, starting with the notion that most systems may not be in equipartition, and we only detect the small fraction that are. This, too, is not a very satisfying scenario because $B$ must then be considerably smaller than \beq\ in hot spots \textit{except} for the ones that we can see, which resemble closer radio galaxies. 

Alternatively, the \hzrgs\ grow so large that the radio hot spots were not correctly associated with host galaxies. We note that the FIRST-detected quasars identified by \cite{volonteri11} are drawn from the \cite{shen11}, who adopt a search radius of 30~arcsec to identify radio counterparts to any SDSS quasars within the FIRST footprint; this radius corresponds to a maximum physical size of $\simlt$260~kpc at $z=$1.4 (above which the size starts to decrease because of cosmology). 
If a large fraction of the high-$z$ radio-loud AGN were significantly larger than $\simgt 500$ kpc, they would have been missed. This hypothesis, too, is far-fetched. 
First, while such large radio galaxies do exist, they represent a small fraction of the nearby population. 
More to the point, for this effect to be at the root of the discrepancy, higher-$z$ radio galaxies ought to be systematically larger than their nearby counterparts, which seems unlikely. It is worth noting that the average density of the Universe at $z\simeq 3$ is comparable to the typical intra-cluster density in a fully virialized, massive cluster at $z\simeq 0.5$, so the overall ambient density of high-$z$ galaxies and their jets is not fundamentally different from their low-$z$ analogs.\\

At the opposite end of the spectrum, most \hzrgs\ could be very compact. Compact steep-spectrum and GHz-peaked sources \citep{ODea1998} are physically small ($<10$~kpc), bright, radio galaxies that may be young systems. Such sources would not be easily quenched by the CMB, but radio galaxies about the same size as their host galaxies would be strongly cooled by their FIR radiation field \citep{smail13}. This scenario implies the existence of spatially resolved analogs at $z\lesssim 0.25$ (where \cxo\ could resolve their lobes), in which dusty starbursts have low radio emission and X-ray bright lobes. We briefly investigated the X-ray and radio emission from several ultraluminous infrared galaxies at low redshift \citep{Tadhunter2011};
these systems span a range of radio morphologies, sizes, and luminosities, but none of them are quenched, and none have notable IC/FIR X-ray emission. This cursory examination is not the final word, but once again demonstrates the challenge of explaining the lack of \hzrgs\ through IC quenching from the starburst.

It also seems unlikely that jets are lower-power, preferentially hadronic, or disrupted, at high redshift. First, as noted in Section~\ref{section:intro}, it is counter-intuitive that the very powerful sources that produce high-$z$ blazars produce very weak jets. Secondly, there is no systematic observational evidence for this (see, however, \citealt{spigola20}, who argue that the bulk Lorentz factor of the $z\simeq 6$ blazar PSO~J030947.49+271757.31 must be relatively low--lower than about 5). \\

Having explored all these possibilities, we are left to scrutinize the definiteness of the claim that the radio-loud quasar population indeed declines substantially at high-$z$ \citep{volonteri11}. 
To start with, it is important to note that their inferences rely on normalizing the number of sources (high-$z$ blazars and luminous SDSS$+$FIRST quasars) by the luminosity functions of their parent populations. This normalization is particularly uncertain when it comes to the highest redshift blazars. \citet{ajello09} derived deconvolved all-sky values based on the number of sources that were observed with an optimal detector, which can lead to non-negligible correction factors. Second, since the Swift-BAT survey found zero blazars at $z>4$, the quoted number of sources in the $4-5$ and $5-6$ redshift bins in \citet{volonteri11} are actually based on the ``minimal evolution" blazar luminosity function proposed by \citet{ghisellini10}, rather than detections. Noting that the luminosity function derived by \citet{ajello09} exceeded the maximum number density of massive black holes allowed by the standard relationships between dark matter halos, galaxies, and their central black holes,  \citet{ghisellini10} imposed an exponential cutoff to the Ajello et al. function above its peak, i.e. at $z=4.3$ (we emphasize that the peak number density itself was poorly constrained, owing to zero blazars detected at $z>4$ and only five at $3<z<4$). 

While a significant population of $z>4$ blazars has been discovered since \citep[e.g.,][and references therein]{Ghisellini2015,Caccianiga2019,belladitta20}, which may still be in tension   with the small number of known $z>4$ radio galaxies, considerable uncertainty remains in the appropriate luminosity function. The \citet{Caccianiga2019} sample is part of a complete, flux-limited sample of radio-selected blazars, for which the number density peaks at $z\simeq 2$, as opposed to $z \simeq 4$ for the \citet{ajello09} sample. The peak at $z \simeq 2$ is consistent with both radio-selected, radio-loud quasars \citep{Mao2017} and radio-quiet quasars \citep{hopkins07}. The volume density of these blazars at $4<z<5$ is $\approx$ 0.1~Gpc$^{-3}$ \citep{Ighina2019}, and since only the few most luminous are BAT sources, this suggests that the expected radio-loud quasar number density at the luminosity threshold adopted by \citet{volonteri11} is substantially lower, thus strongly reducing the tension between the number of blazars and radio-loud quasars at $z>4$. However, it is not clear whether the discrepancy with \citet{ajello09} results from the radio and X-ray selection functions sampling very different objects, or if the most luminous blazars, seen with BAT and tracing the most massive black holes, evolve differently.

Meanwhile, an updated quasar bolometric luminosity function \citep{shen20} revises the normalization downwards relative to \citet{hopkins07}, and has a steeper slope at $\simgt 10^{46}$~erg~s$^{-1}$ and $z\simgt 2$. For a given radio-loud fraction, this would tend to worsen the discrepancy highlighted by \citet{volonteri11}, but may also require a revision of the \citet{ghisellini10} minimal evolution luminosity function.

In addition to the uncertainty in the luminosity functions, the selection of radio-loud galaxies from the SDSS$+$FIRST sample carries uncertainty from the quasar selection algorithm. For example, \citet{volonteri11} adopted the \citet{Richards2002} SDSS color selection algorithm to identify extremely luminous ($\simgt 10^{47}$~erg~s$^{-1}$), radio-loud quasars, whereas, e.g., \citet{Mao2017} used a variety of data, including optical spectra and infrared colors to identify SDSS$+$FIRST radio-loud quasars. As \citet{volonteri11} point out, incompleteness bias should cancel out when measuring the radio-loud fraction in an incomplete sample, but a bias for or against classifying quasars as radio-loud with a given method \citep[e.g., due to a lack of infrared photometry;][]{Banados2015} would still impact the measurement. This issue is compounded by the small number of radio-loud quasars identified by \citet{volonteri11} at $z \simgt 4$, where the statistical uncertainty and cosmic variance could easily change the reported fractions by several percent. Indeed, \citet{Banados2015} find no evolution in the radio-loud fraction up to $z \simgt 6$.\\

To summarize, we have failed to identify a viable explanation for why the fraction of radio-loud quasars would drop substantially above $z \simgt 3$. Specifically, we demonstrate that IC/CMB can quench the radio lobes of high-$z$ jetted AGN, but up to $z \simgt 6$ it cannot quench the radio emission from their hot spots. On the other hand, we argue that the reported tension between the observed and expected radio-loud fraction is substantially uncertain because the luminous blazar luminosity function is uncertain at $z \simgt 3$, and because there is uncertainty in the classification of quasars as radio-loud or radio-quiet, with a magnitude that depends on the available data. Thus, both the expected number of radio-loud quasars based on the blazar luminosity function and the expected number extrapolated from the measured radio-loud fraction are probably known imprecisely. The true magnitude of the tension, if any, remains unclear.

\section{Conclusions}

We have undertaken a critical, definitive assessment of the role of IC scattering of CMB photons in the context of radio galaxies. At low-$z$, cooling of relativistic particles in the magnetized, jet-powered lobes and hot spots is typically dominated by radio synchrotron emission, and IC/CMB is negligible. Because of the steep, $\propto (1+z)^4$ dependence of the CMB radiation energy density, IC/CMB is supposed to become progressively more important with respect to synchrotron cooling as the redshift increases. For typical energies at play, this process will boost CMB photons into the X-ray band, and is thus expected to yield a $z$-dependent, concurrent X-ray brightening and radio dimming of the extended jet-powered structures. Yet, observational evidence for this seemingly unavoidable effect so far remains sparse and controversial. 

Here we show that high-resolution radio and X-ray imaging data, where the emission from the hot spots can be separated out from the lobes, are necessary to properly assess and quantify the role of IC/CMB. This is because, in addition to redshift, the expected X-ray luminosity arising from IC/CMB depends on the strength of the magnetic field. Under the assumption of equipartition, the latter can be expressed in terms of the radio luminosity and size of the emitting region (Equation~\ref{eqn:beq}).
Accordingly, whereas IC/CMB can be expected to play a significant role in cooling the extended, weakly dominated lobes, it will be more easily offset by synchrotron and SSC cooling in the compact, highly magnetized hot spots.

Analysing spatially resolved radio and X-ray data for sample of 11 high-$z$ radio galaxies ($1.3\simlt z \simlt 4.3$) we demonstrate that:

\begin{itemize}
\item{X-ray emission from the lobes is fully consistent with the expectations from IC/CMB in equipartition (Figure \ref{fig:lx_lr}). This is true regardless of redshift, AGN luminosity, and lobe size (Figure \ref{fig:b_eq}). }
\item 
{In contrast, X-ray emission from the hot spots of the \hzrgs\ in our sample is consistent with the expectation from SSC in equipartition (Figure \ref{fig:hardcastle}).  }
\item 
{The galaxy sample examined in this work is likely to be both the high-$z$ analog of local radio galaxies, as well as a fair a representation of the broader population of \hzrgs, in that they appear to be close to equipartition over a wide range in power and size. Although this sample does not include radio galaxies up to the highest redshifts to which blazars have been detected, the \hzrgs\ in this sample are consistent with being the misaligned counterparts of the luminous blazars studied in \citet{ghisellini10}.}
\item 
{Once the dependence on size and radio luminosity are property accounted for, the measured lobe X-ray luminosity does indeed bear the characteristic $\propto (1+z)^4$ dependence expected of a CMB seed radiation field (Figure \ref{fig:lxlr_z}).}

\item 
{IC/CMB causes a concurrent, $z$-dependent X-ray brightening/radio dimming of the lobes of radio galaxies. Whereas this effect can quench the radio emission from the lobes of \hzrgs\ (above $z\simgt 3$) below the sensitivity threshold of the FIRST survey, it would not be sufficient to quench the radio emission from their strongly magnetized hot spots at $z \simlt 6$.
Thus, IC/CMB alone can not be responsible for a deficit in high-$z$, radio-loud AGN if--as we argue--such AGN typically have bright hot spots.}

\end{itemize}

\section*{Data Availability}

The data used in this work are publicly available through the VLA data archive (\url{https://science.nrao.edu/facilities/vla/archive/index}) and the Chandra data archive (\url{https://cxc.cfa.harvard.edu/cda/}), using the project codes and ObsIDs from Table~\ref{table:sample}. Where necessary, the data were processed using standard tools for each telescope available from those facilities.

\section*{Acknowledgments}

We thank the anonymous reviewer for catching mistakes and giving comments that improved this manuscript. 

EG was partially suppported by the National Aeronautics and Space Administration through Chandra Award Number G06-17082X issued by the Chandra X-ray Center (CXC), which is operated by the Smithsonian Astrophysical Observatory for and on behalf of the National Aeronautics Space Administration under contract NAS8-03060. 

The scientific results reported in this article are based to a significant degree on data obtained from the Chandra Data Archive. This research has made use of software provided by the CXC in the CIAO application package.

The National Radio Astronomy Observatory is a facility of the National Science Foundation operated under cooperative agreement by Associated Universities, Inc.

This research has made use of the NASA/IPAC Extragalactic Database (NED),
which is operated by the Jet Propulsion Laboratory, California Institute of Technology, under contract with the National Aeronautics and Space Administration.

\bibliographystyle{mnras}

\end{document}